\documentclass[extra,mreferee]{gji}
\usepackage{timet}
\usepackage{amsfonts,amssymb,amsmath,graphpap}
\usepackage{graphicx}
\usepackage{color}
\usepackage{multirow}
\providecommand\be{\begin{equation}}
\providecommand\ee{\end{equation}}
\providecommand\bsub{\begin{subequations}}
\providecommand\esub{\end{subequations}}

\providecommand\bcdot{\mbox{\boldmath $\cdot$}}

\providecommand\bnabla{\mbox{\boldmath $\nabla$}}
\newcommand{\ve}[1]{\mbox{\boldmath $#1$}}

\title[ANPAR]
  {Analytical Parameterization of Self-Consistent
Polycrystal Mechanics: Fast Calculation of Upper Mantle Anisotropy}
\author[Neil J. Goulding et al.]
  {Neil J. Goulding$^1$\thanks{Corresponding author (ng12245@bristol.ac.uk)}, Neil M. Ribe$^2$\thanks{ (ribe@fast.u-psud.fr)}, Olivier Castelnau$^3$, Andrew M. Walker$^4$, James Wookey$^1$\\
  $^1$ School of Earth Sciences, University of Bristol, Wills Memorial Building, Queen's Road, Bristol BS8 1RJ, UK.\\
$^2$ Lab FAST, Univ Paris-Sud/CNRS, Bat 502, Campus Univ, Orsay, F-91405, France.\\
$^3$ Lab PIMM, CNRS, Arts et M\'{e}tiers ParisTech, 151 Bd de l'Hopital, 75013 Paris, France\\
$^4$ School of Earth and Environment, University of Leeds, Leeds, LS2 9JT, UK.
  }
\date{}

\begin{document}

\label{firstpage}

\maketitle

\begin{summary}

 Progressive deformation of upper mantle rocks via dislocation creep
causes their constituent
crystals to take on a non-random orientation distribution
(crystallographic preferred orientation or CPO) whose observable
signatures include shear-wave splitting and azimuthal
dependence of surface wave speeds. Comparison of these signatures with 
mantle flow models thus allows mantle dynamics to be unraveled on global and regional scales. However, existing self-consistent models of CPO evolution are computationally expensive when
used in 3-D and/or time-dependent convection models. Here we propose a new
method, called ANPAR, which is based on an analytical parameterisation 
of the crystallographic spin predicted by the second-order (SO)
self-consistent theory. Our parameterisation
runs $\approx 2$-3 $\times 10^4$ times faster than the SO model
and fits its predictions for CPO and crystallographic spin with a variance reduction $>$ 99\%. We 
illustrate the ANPAR model predictions for three uniform deformations (uniaxial compression, pure shear, simple shear) and for a corner-flow
model of a spreading ridge. 
\end{summary}


\section{Introduction}

Seismic anisotropy observed in Earth's upper mantle is typically explained by the partial alignment of the lattices of the constituent olivine and pyroxene crystals caused by deformation associated with mantle convection (e.g. Nicolas and Christensen, 1987; Silver, 1996; Long \& Becker 2010). Because each crystal is elastically anisotropic, this non-random distribution of crystallographic directions (called a crystallographic preferred orientation, or CPO) will impart elastic anisotropy to the bulk material. The seismically observable consequences of this anisotropy include shear-wave birefringence or `splitting' (e.g. Crampin, 1984; Silver and Chan, 1991) and the azimuthal dependence of surface-wave speeds (e.g. Montagner \& Tanimoto, 1991). Simulation of the development of CPO in models of mantle deformation, and comparison of this with seismic observations of the Earth, allow mantle dynamics to be unraveled on global (e.g. Becker et al., 2012) and regional scales (e.g. Long, 2013). However, these simulations are computationally challenging when performed for time-dependent models of mantle convection or at the high spatial resolution needed for finite frequency simulation of seismic wave propagation. Here we describe an accurate but computationally efficient alternative to existing methods for the simulation of CPO development in the upper mantle.

The principal cause of CPO and seismic anisotropy in the mantle is the progressive deformation experienced by mantle rocks as they participate in the global convective circulation. Under appropriate conditions of stress, temperature, and grain size, olivine and pyroxene crystals deform via dislocation creep, whereby internal dislocations move through the crystal to accommodate strain. The dislocations move on crystallographic planes and in directions set by the crystal structure, and the combination of a plane and direction define the limited number of slip systems available to allow the crystal to deform. Deformation of this type constrains the crystallographic axes to rotate relative to a fixed external reference frame, much as a tilted row of books on a shelf rotates when one pushes down on it. Because crystals with different orientations rotate at different rates, the overall distribution of orientations evolves with time in a way that reflects both the geometry of the slip systems and the character of the imposed deformation.

Because CPO and seismic anisotropy are so directly linked to
progressive deformation,  observations of seismic anisotropy
have the potential to constrain the pattern of convective
flow in the mantle. Realizing this potential, however,
requires a reliable polycrystal mechanics model that
can predict how the individual crystals in an aggregate deform
and rotate in response to an imposed macroscopic
stress or strain rate. Three broad classes of polycrystal
models have been proposed to date. 

The first class comprises the “full-field” models. In these, the polycrystal is treated explicitly as a spatially extended body, and the stress and strain within it are field variables that vary continuously as a function of position. Full-field models allow the stress and strain to vary both among and within individual grains in a physically realistic way. This approach can be implemented as a finite element problem (e.g. Sarma \& Dawson, 1996; Kanit \textit{et al}., 2003) or, more efficiently, using a method based on Fast Fourier Transforms (Moulinec \& Suquet, 1998; Lebensohn, 2001; Suquet \textit{et al}., 2012). Predictions from full-filed models agree remarkably well with laboratory experiments (Grennerat \textit{et al}., 2012) and analytical results available for simple cases (Lebensohn \textit{et al}., 2011). However, their great computational expense makes them too slow (by many orders of magnitude) for routine use in convection calculations. 

This disadvantage is overcome to some extent by  so-called `homogenisation' models, in which the detailed spatial distribution of the grains
is ignored and the aggregate is treated as a finite number of grains with different orientations and material
properties. In this mean-field approach compatibility of stress and strain equilibrium is not enforced between
spatially contiguous grains, but rather between each grain
and a `homogeneous effective medium' defined by the average of all the other
grains. The best-known member of this class of models is based on the viscoplastic self-consistent (VPSC) formalism (Molinari \textit{et al.} 1987, Lebensohn \& Tom\'e 1993), in which
the local
stress and strain rate tensors vary among the
grains. The VPSC model has been widely used in solid-earth geophysics including studies of CPO development in the upper mantle (e.g. Wenk \textit{et al.}, 1991, 1999; Tommasi \textit{et al.} 1999, 2000, 2009; Mainprice \textit{et al.}, 2005; Bonnin \textit{et al.}, 2012; Di Leo \textit{et al.}, 2014), the transition zone (Tommasi \textit{et al.} 2004) and the lowermost mantle (Merkel \textit{et al.}, 2007; Wenk \textit{et al.}, 2006, 2011; Mainprice \textit{et al.}, 2008; Walker \textit{et al.}, 2011; Dobson \textit{et al.}, 2014, Nowacki \textit{et al.}, 2013; Amman \textit{et al.}, 2014; Cottaar \textit{et al.}, 2014). 
However, as noted by Masson \textit{et al}. (2000), the VPSC model suffers from a theoretical inconsistency
in the definition of the stress localisation tensor.

More recently, an improved `second order' (SO) version of the homogenisation scheme
has been proposed by Ponte Casta\~neda (2002).  
In the SO model the stress and strain rate varies among grains with the same 
orientation and physical properties. As a result, its predictions of quantities such as the
effective average stress in the aggregate are much more
accurate than those of simpler homogenisation schemes
such as VPSC (Castelnau \textit{et al}. 2008). 
Recent examples of the application of the SO approach to olivine deformation are provided by Castelnau \textit{et al}. (2008, 2009, 2010) and Raterron \textit{et al}. (2014).

While the physical self-consistency of the SO  
and (to a lesser extent) VPSC models is appealing, 
both are computationally expensive when applied to
typical mantle minerals deforming by dislocation creep.
The reason is the strongly nonlinear rheology of such minerals,
which makes it necessary to use iterative methods
to solve the equations of stress compatibility among the large number
($\sim 10^3-10^4$) of grains required to represent the polycrystal.
Moreover, the number of iterations required at each deformation step
increases rapidly as the CPO becomes progressively more strongly anisotropic. 
These difficulties render the VPSC and SO models unsuitable for 
calculations of evolving CPO in complex time-dependent mantle
flow fields, unless powerful computer capacity and elaborate computation strategy are used. Indeed, because of these computational constraints, none of the studies referenced above make use of the VPSC or SO approaches to directly compute the elasticity on a fine spatial scale (suitable for finite frequency forward modelling of the seismic wave field) from a time-varying description of mantle flow. Instead various approximations are used, such as limiting the calculation to selected ray-theoretical paths (Di Leo \textit{et al.} 2014, Nowacki \textit{et al.} 2013), interpolating the calculated elasticity (Bonnin \textit{et al.} 2012), or simplifying the model of mantle flow (Raterron \textit{et al}. 2014).

A final degree of physical simplicity and computational efficiency is reached in models of the `kinematic' class, which are based on either an analytical expression for the deformation-induced rate of crystallographic rotation (Ribe \& Yu, 1991; Kaminski \& Ribe, 2001; Kaminski \textit{et al}., 2004) or on a simple relationship between finite strain and the expected CPO (Muhlhaus \textit{et al}. 2004; Lev \& Hager, 2008). One example, the DRex model (Kaminski and Ribe, 2001; Kaminski et al., 2004) has been widely used to predict CPO and seismic anisotropy from flow models (e.g., Lassak \textit{et al}., 2006; Conder \& Wiens, 2007; Becker, 2008; Long \& Becker, 2010; Faccenda \& Capitanio, 2012, 2013; Faccenda, 2014).
Kinematic models are computationally 10-100 times faster than homogenisation
models, and predict very similar CPO.
However, the physical principle underlying the expression for the spin
is \textit{ad hoc}, and has not yet been adequately justified. Moreover, 
because the kinematic approach does not account explicitly for stress compatibility among
grains, it cannot be used to predict rheological properties of a deforming aggregate.

In view of the above limitations, it would clearly be desirable to have a polycrystal model 
that combines the physical rigor of the self-consistent approach with a 
much lower computational cost. The aim of this paper is to 
derive such a model. For purposes of illustration, we consider the case of
a pure olivine polycrystal (dunite), a relevant (albeit simplified)
representation of the mineralogy of the upper $\approx 400$ km of Earth's mantle.  
Our approach is to examine in detail the predictions of the SO model for dunites subject to 
different kinds of deformation, and to extract from those predictions a
simple parameterisation that can be expressed analytically. 

The most important prediction of the SO model is the 
crystallographic spin $\dot{\ve g}$ as a function of the crystal's 
orientation $\ve g$, which is what controls the evolution of CPO.
Accordingly, this paper focusses on the task of finding an
analytical parameterisation of $\dot{\ve g}(\ve g)$ that 
agrees with the SO model predictions. We first note that the total spin $\dot{\ve g}$ is 
the sum of spins $\dot{\ve g}^{[s]}$ due to the activities of each of the 
slip systems $s = 1, 2, ..., S$ within the crystals. We then derive an analytical 
expression for $\dot{\ve g}^{[s]}(\ve g)$ for the limiting case of an
aggregate of crystals with only a single active slip system ($S=1$). 
This expression is then compared, for each slip system separately,
with the spins $\dot{\ve g}^{[s]}(\ve g)$ predicted by the SO model
for an aggregate of crystals with several simultaneously active slip systems  
($S > 1$). Remarkably, we find
that the analytical expression for $\dot{\ve g}^{[s]}(\ve g)$ matches
the SO prediction exactly for each slip system $s$, to within a set of
amplitudes $A_{ijkl}$ that can be determined by least-squares 
fitting. We uncover surprising symmetries that reduce the number of independent non-zero components of the `spin' tensor $\mathbf{A}$ from 25 to just 2. Finally, we use full SO solutions to determine how these
coefficients depend on the relative strengths of the 
slip systems and on the finite strain experienced by the aggregate. 

For irrotational, time-independent deformation, the finite strain ellipsoid has the same shape and orientation as the virtual ellipsoid generated by the instantaneous global strain-rate tensor. In this simplified case we show that we require only one amplitude. However, when the two ellipsoids are not aligned (see Fig. 1), an extra amplitude is required.
We show that predictions of evolving CPO using these analytical parameterisations (which we call ANPAR) are
indistinguishable from those of the SO model, and cost only
$\approx 0.01\%$ as much time to compute.  

\begin{figure}\begin{center}\scalebox{0.4}{\includegraphics{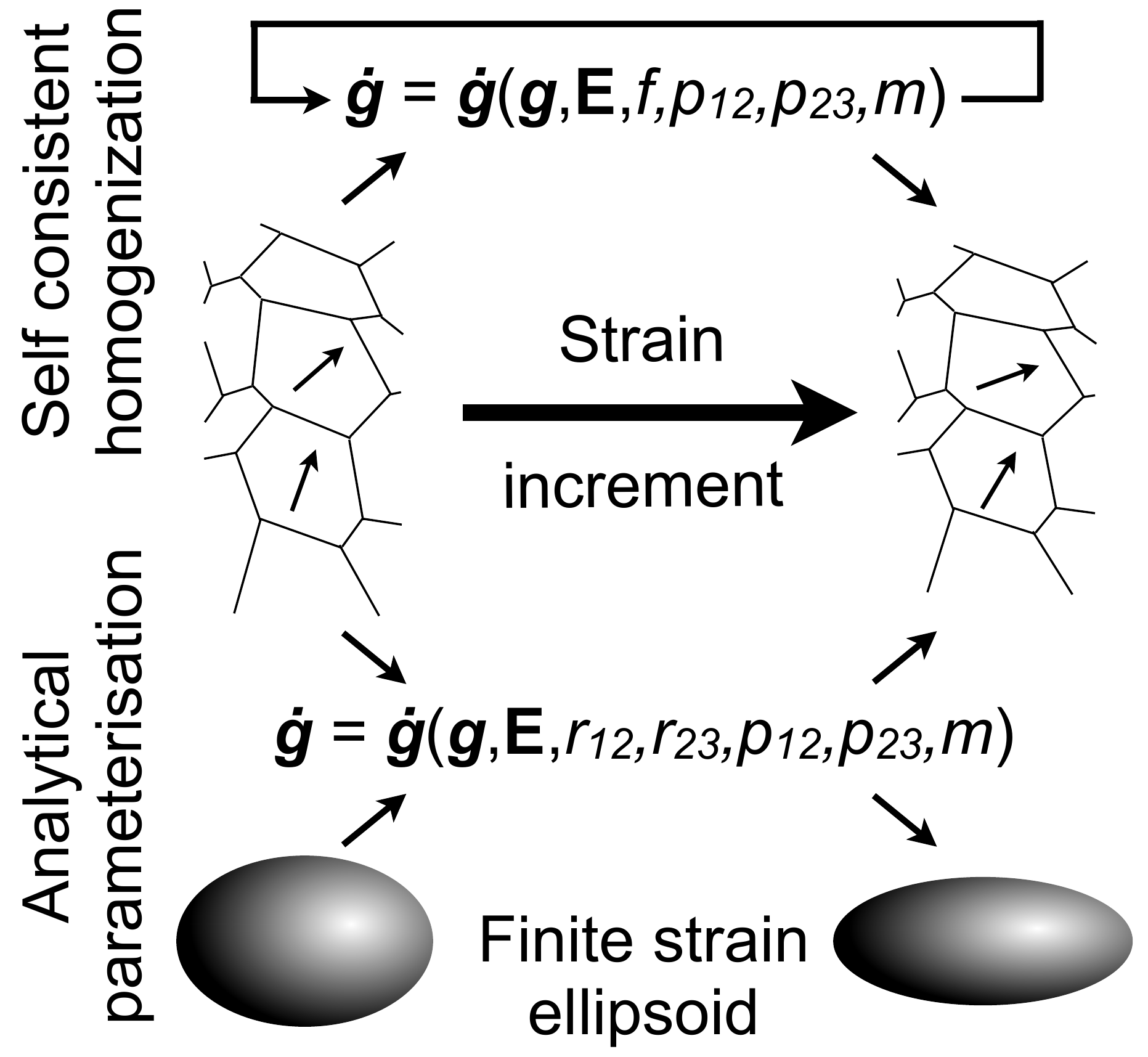}}\end{center}
\caption{\label{Fig1}Schematic comparison of the analytical approach and homogenisation methods for a single strain increment. Using SO to calculate the spin, $\dot{\ve g}$, for the $n^{\textrm{th}}$ grain and update its orientation, $\ve g$, requires knowledge of the strain, and thus spin, of all other crystals in the aggregate necessitating an expensive self-consistent solution. Our analytical approach replaces this information with a record of previous deformation stored as an auxiliary finite strain ellipsoid. This, combined with a handful of other parameters, $A$, enables rapid calculation of the spin.}\end{figure}

\section{Theoretical preliminaries}

We begin by reviewing how the orientation and internal deformation
of crystals in an aggregate
are described mathematically, using the particular
case of olivine as an example.

\subsection{Crystal orientation and orientation distribution}
\label{orientation}

Consider an aggregate comprising a 
large number $N$ of olivine crystals deforming by dislocation creep.  
When the aggregate as a whole is subject
to a given macroscopic deformation, its constituent crystals respond by
deforming via internal shear on a small number $S$ of `slip systems'.
Each slip system $s = 1, 2, ... , S$ is defined
by a unit vector $\ve n^{[s]}$ normal to the slip (glide) plane and a unit (Burgers) vector
$\ve l^{[s]}$ parallel to the slip direction. In this study we assume that
olivine has three dominant slip systems
(010)[100], (001)[100] and (010)[001], corresponding to the indices
$s = 1$, 2, and 3, respectively.

\begin{figure}
\begin{center}
\includegraphics[scale=0.8]{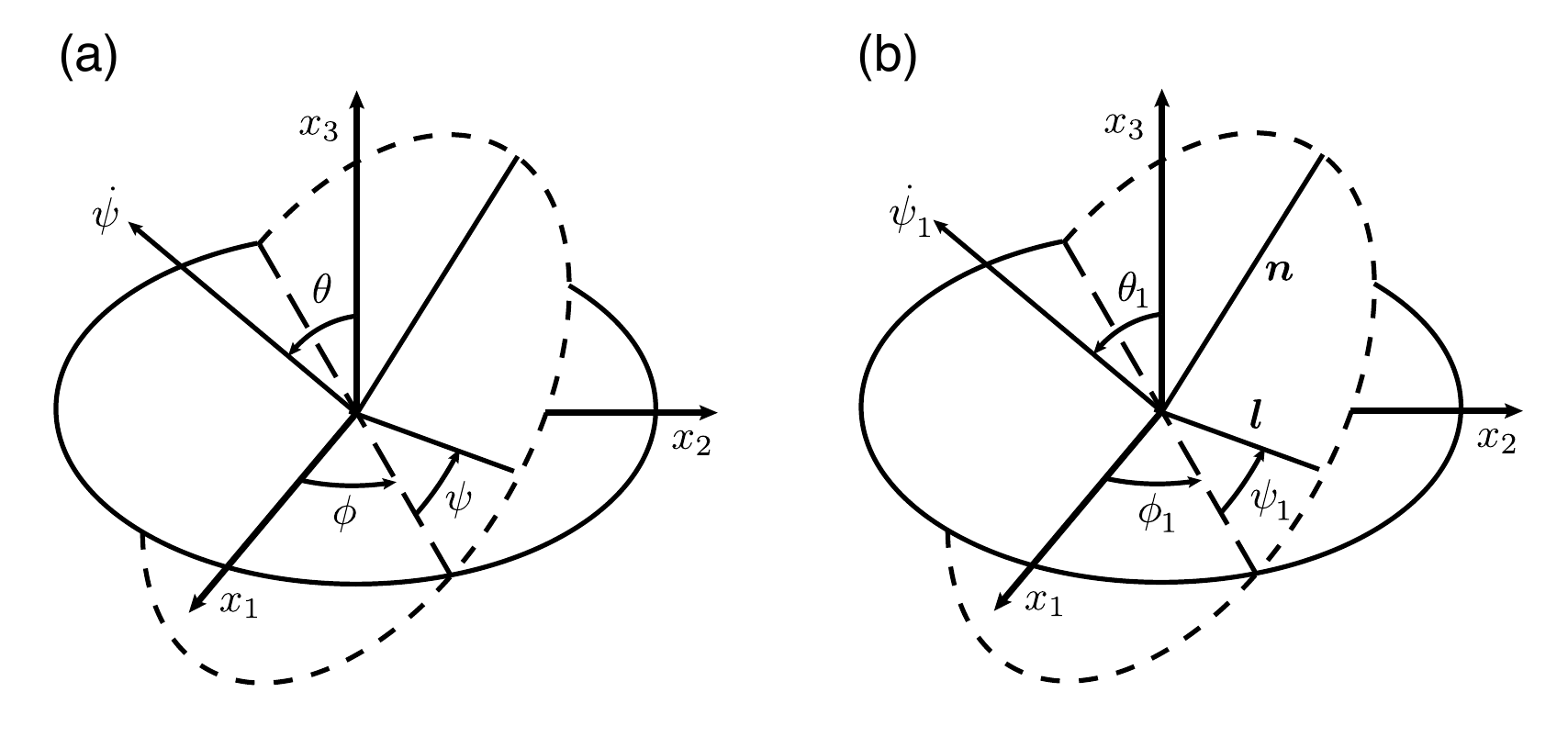}
\end{center}
\caption{\label{fig_eulerangles}
(a) General definition of the Eulerian angles $(\phi, \theta, \psi)$. (b) Definition
of slip-system specific Eulerian angles $(\phi^{[s]}, \theta^{[s]}, \psi^{[s]})$ such that
both the slip direction $\ve l^{[s]}$ and the normal $\ve n^{[s]}$ to the slip plane are
in the plane perpendicular to the $\dot{\psi}^{[s]}$-axis. 
}
\end{figure}

The degree of anisotropy of an aggregate can be
described by specifying for each crystal the three Eulerian
angles $(\phi, \theta, \psi)\equiv \ve g$ that describe its orientation
relative to fixed external axes. The definition of these angles that
we use is shown in Fig. \ref{fig_eulerangles}. The associated matrix
of direction cosines $a_{ij}$ is 
\be
a_{ij}(\ve g) =
\left(
\begin{array}{ccc}
$$\mathrm{c}\phi\,\mathrm{c}\psi - \mathrm{s}\phi\,\mathrm{s}\psi\,\mathrm{c}\theta$$  &  $$\mathrm{s}\phi\,\mathrm{c}\psi + \mathrm{c}\phi\,\mathrm{s}\psi\,\mathrm{c}\theta$$ &  $$\mathrm{s}\psi\,\mathrm{s}\theta$$\\
$$- \mathrm{c}\phi\,\mathrm{s}\psi - \mathrm{s}\phi\,\mathrm{c}\psi\,\mathrm{c}\theta$$  &  $$-\mathrm{s}\phi\,\mathrm{s}\psi + \mathrm{c}\phi\,\mathrm{c}\psi\,\mathrm{c}\theta$$ & $$\mathrm{c}\psi\,\mathrm{s}\theta$$\\
$$\mathrm{s}\phi\,\mathrm{s}\theta$$  &  $$- \mathrm{c}\phi\,\mathrm{s}\theta$$ &  $$\mathrm{c}\theta$$\\
\end{array}
\right),
\label{aij}
\ee
where $\mathrm{c}$ and $\mathrm{s}$ indicate the cosine and sine, respectively,
of the angle immediately following. The quantity $a_{ij}$ is
the cosine of the angle between the crystallographic
axis $i$ and the external axis $j$. 

In the limit as the number of grains $N\rightarrow\infty$, the
distribution of their orientations can be described by a continuous
`orientation distribution function' (ODF) $f(\ve g,t)$,
defined such that $f(\ve g,t) \mathrm{d}\ve g$ is the volume fraction of crystals
with orientations between $\ve g$ and $\mathrm{d}\ve g$ at time $t$.
For crystals with triclinic symmetry, the volume of the space of
Eulerian angles (`Euler space') required to include all possible
orientations is $\phi\in [0, 2\pi]$,
$\theta\in [0, \pi]$, $\psi\in [0, 2\pi]$. For olivine, which is orthorhombic,
the required volume of the Euler space is reduced to $\phi\in [0, \pi]$,
$\theta\in [0, \pi]$, $\psi\in [0, \pi]$. The condition that the total volume fraction of crystals
with all possible orientations is unity is then
\be
\int f(g,t)\mathrm{d}g\equiv
\int_0^{\pi}\int_0^{\pi}\int_0^{\pi}f(g,t)\;\mathrm{d}\phi\; \mathrm{d}\psi \;\mathrm{d}\theta
\,\sin\theta\,  = 1,
\label{continuity}
\ee
which implies that $f = (2\pi^2)^{-1}\equiv f_0$
for an isotropic orientation distribution. 

\subsection{Kinematics of intracrystalline slip}

The time evolution of the ODF
is governed by the equation (Clement 1982)
\be
0 = \frac{\partial f}{\partial t} + \bnabla\bcdot (\dot{\ve g} f) \equiv
\frac{\partial f}{\partial t}
+ \frac{\partial}{\partial\phi}(\dot{\phi} f)
+ \frac{\partial}{\partial\psi}(\dot{\psi} f)
+ \frac{1}{\sin\theta}\frac{\partial}{\partial\theta}(\dot{\theta} \sin\theta f),
\label{evoleqn}
\ee
where $(\dot{\phi}, \dot{\theta}, \dot{\psi})\equiv \dot{\ve g}$ is the rate of change of
the orientation (`spin') of an individual crystal with orientation $\ve g$. Eqn. (\ref{evoleqn}) is a conservation
law which states that
the rate of change of the volume fraction of crystals having orientations
in a small element $\mathrm{d}\ve g$ of the Euler space is equal to the net flux of crystal orientations
into that element. The spins $(\dot{\phi}, \dot{\theta}, \dot{\psi})$ are related to the 
Cartesian components $\omega_i$ of the spin by
\bsub
\label{gdotdef}
\be
\dot\phi = \omega_3 + (\omega_2 \cos\phi - \omega_1\sin\phi)\cot\theta ,
\ee
\be
\dot\theta= \omega_1 \cos\phi + \omega_2 \sin\phi,
\ee
\be
\dot\psi =  (\omega_1 \sin\phi - \omega_2\cos\phi)\csc\theta.
\ee
\esub
Note also that the crystallographic spin $\boldsymbol\omega$ is just
the sum of the externally imposed rotation rate $\boldsymbol{\Omega}$
and a contribution $\boldsymbol{\omega}^{(c)}$ due to intracrystalline slip, or 
\be
\omega_i = \Omega_i - \epsilon_{ijk} l_j n_k\dot\gamma
\equiv  \Omega_i + \omega_i^{(c)}.
\label{omega}
\ee

The spin $\dot{\ve g}$ is the fundamental quantity that will concern us in this study. 
It depends on the instantaneous macroscopic velocity gradient tensor
$\mathbf D$, the components of which are
\be
D_{ij} = E_{ij} - \epsilon_{ijk}\Omega_k,
\label{bigdij}
\ee
where $E_{ij}$ and $\Omega_k$ are the components of the strain rate tensor $\mathbf E$ and
the macroscopic rotation rate $\mathbf{\Omega}$ of the polycrystal, respectively.

When the aggregate is  deformed, each
crystal within it responds by deforming in simple shear on planes
normal to $\ve n(\ve g)$ at a rate $\dot\gamma(\ve g)$. 
The local velocity gradient tensor inside the crystal is thus
\be
d_{ij} = \dot\gamma l_i n_j.
\label{velgrad}
\ee
The local strain rate tensor $e_{ij}$ is the symmetric part of $d_{ij}$, or
\be
e_{ij} = \frac{\dot\gamma}{2}(l_i n_j + l_j n_i) \equiv \dot\gamma S_{ij}.
\ee
Here $S_{ij}$ is the Schmid tensor, which resolves the strain-rate inside each crystal onto the natural frame of reference of the slip system. It is symmetric and traceless and therefore has only five independent components. 
These can be expressed in terms of 
generalized spherical harmonics $T_l^{mn}$ of degree $l=2$, where
\be T_l^{m n}(\phi,\theta,\psi)
=e^{i m\psi}P_l^{m n}(\cos\theta)e^{i n\phi},
\ee
and $P_l^{m n}(\cos\theta)$ is the associated Legendre polynomial (Bunge 1982, eqn. 14.2).
Explicit expressions for the independent components of the (slip-system specific) Schmid
tensor $S^{[s]}_{ij}$ for 
the slip systems $s = 1, 2, 3$, are given in Appendix A.

Another kinematical object that plays an important role in our theory is the
finite strain ellipsoid (FSE) associated with the deformation history
experienced by a polycrystal. It is well known in fluid mechanics that
an arbitrary time-dependent flow field transforms an initially spherical 
fluid element of infinitesimal size into an ellipsoid, called the FSE.
The shape of the FSE can be characterized by 
the logarithms of the ratios of the lengths $c_1$, $c_2$ and $c_3$ of its axes, 
viz. 
\be
r_{12} = \ln\frac{c_1}{c_2},
\quad
r_{23} = \ln\frac{c_2}{c_3},
\quad
r_{31} = \ln\frac{c_3}{c_1}.
\label{r12r23}
\ee
Incompressibility of the fluid implies $r_{12} + r_{23} + r_{31} = 0$,
so that only two of the quantities $r_{ij}$ are independent.
We also define an `equivalent strain'
\be
r_0=\frac{\sqrt{2}}{3}\left(
r_{12}^2 + r_{23}^2+ r_{31}^2
\right)^{1/2}
 = \frac{2}{3}
\left(
r_{12}^2 + r_{12} r_{23} + r_{23}^2
\right)^{1/2}
\label{r0}
\ee

\subsection{Slip-system rheology}

Following standard practice, we assume that 
the slip rate $\dot{\gamma}^{[s]}$ on
each slip system $s$ obeys a power-law rheology of
the form
\be
\dot{\gamma}^{[s]}\propto
\left|\frac{\tau}{\tau^{[s]}}\right|^{m^{[s]} - 1}
\frac{\tau}{\tau^{[s]}},
\ee
where $\tau$ is the resolved shear stress (i.e., the shear stress
acting on the slip plane in the slip direction), $\tau^{[s]}$ is a
`critical resolved shear stress' (CRSS) that measures the inherent
resistance of the slip system to slip, and $m^{[s]}$ is a power-law exponent. Although the standard notation is to use $n$ as the stress exponent, we have chosen $m$ in this paper to avoid confusion with all the different occurences of $n$. We assume $m^{[s]} = 3.5$ for all slip systems, following Bai \textit{et al.} (1991).
Because the macroscopic deformation rate
of the aggregate is specified in our SO calculations, only the ratios
of the parameters $\tau^{[s]}$ (and not their absolute values) are relevant.
In our calculations we assume
$\tau^{[1]}/\tau^{[2]}\in [0.25, 4.0]$ and $\tau^{[2]}/\tau^{[3]}\in [0.25, 4.0]$ (see Table 1).
\begin{table}
\label{tab-scales}
\caption{Slip systems}
\begin{tabular}{ccccc}
\hline
\multicolumn{1}{c}{index $s$}  &
\multicolumn{1}{c}{slip plane}  &
\multicolumn{1}{c}{slip direction}  &
\multicolumn{1}{c} {$\tau^{[s]}/\tau^{[2]}$} &
\multicolumn{1}{c}{exponent} \\
\hline
1 & (010) & [100] &  0.25-4.0  & 3.5  \\
2 & (001) & [100] &  1.0   &  3.5  \\
3 & (010)  & [001] &  0.25-4.0   &  3.5 \\
4 & (101)  & [$\overline{1}$01] &  100.0   &  3.5 \\
5 & (10$\overline{1}$)  & [101] &  100.0   &  3.5 \\
\end{tabular}
\end{table}
We characterize the CRSS ratios of the dominant
slip systems $s=1$, 2 and 3 in terms of the variables
\be
p_{12} = \ln\frac{\tau^{[1]}}{\tau^{[2]}},
\;\;\;\;\;\;
p_{23} = \ln\frac{\tau^{[2]}}{\tau^{[3]}}.
\label{p123}
\ee
Note also that
\be
p_{31} = \ln\frac{\tau^{[3]}}{\tau^{[1]}}=- p_{12} - p_{23}.
\label{p13}
\ee

The SO model requires that each crystal in the aggregate satisify
von Mises's criterion, according to which a crystal can only accomodate
an arbitrary imposed deformation if it has at least five independent slip systems. 
This is not the case for olivine, and so to ensure numerical convergence 
we assume that each olivine crystal has, in addition to the 
three dominant slip systems mentioned previously, two 
harder systems, namely $(101)[\overline{1}01]$ and $(10\overline{1})[101]$. In our calculations we assume
$\tau^{[4]}/\tau^{[2]} = \tau^{[5]}/\tau^{[2]} = 100$ (see Table 1).
While these slip systems contribute significantly to the intracrystalline
stress, they have a negligible ($\approx 1\%$) effect on the slip rates on the dominant
systems. The model therefore gives valid predictions of the evolution
of CPO. 

\section{Analytical parameterisation}

The considerations of the previous section imply that the instantaneous crystallographic 
rotation rate $\dot{\ve g}$ depends on the crystal's orientation $\ve g$; the 
macroscopic strain rate tensor $\mathbf{E}$;  the already existing texture $f$;
and the parameters $p_{12}$,
$p_{23}$ and $m$ that characterize the rheology of the slip systems: 
\be
\dot{\ve g}=\dot{\ve g}\left(\ve g,\mathbf{E}, f, p_{12}, p_{23}, m\right).
\label{gdot1}
\ee
Here $\dot{\ve g}$ is understood as the slip-induced rotation rate, without
the contribution due to the macroscopic vorticity which is the same for 
all crystals and can simply be added to $\dot{\ve g}$.

Next, we note that the 
spin components (\ref{gdotdef}) take a particularly simple form
when rewritten in terms of slip system-specific 
Eulerian angles $(\phi^{[s]}, \theta^{[s]}, \psi^{[s]})$ defined so
that both the slip vector $\ve l^{[s]}$ and the vector $\ve n^{[s]}$ normal to the slip
plane are perpendicular
to the $\dot{\psi}^{[s]}$-axis (Fig. \ref{fig_eulerangles}b). The 
crystallographic spin $\dot{\ve g}^{[s]}$ produced by slip in
the direction $\ve l^{[s]}$ on the plane $\ve n^{[s]}$ then has only 
a single non-zero component
$\dot\psi^{[s]}$, and $\dot\phi^{[s]} = \dot\theta^{[s]} = 0$ identically. 
Fig. \ref{fig_eulerangles}b implies that
$l^{[1]}_i = a_{1i}$ and $n^{[1]}_i = a_{2i}$, where
$a_{ij}$ are given by (\ref{aij}) with 
$(\phi, \theta, \psi) \to (\phi^{[s]}, \theta^{[s]}, \psi^{[s]})$. 
Ignoring the macroscopic vorticity as explained above, we find that  
(\ref{gdotdef}) and (\ref{omega}) simplify to
\be
\dot{\phi}^{[s]} = \dot{\theta}^{[s]} = 0,
\;\;\;\;\dot{\psi}^{[s]}=-\dot\gamma^{[s]}({\ve g^{[s]}}).
\label{psidoteqgammadot}\ee
Thus the crystallographic spin due to 
slip is simply the negative of the shear rate 
on the slip system in question. 

\begin{figure}
\begin{center}\scalebox{0.5}{
\includegraphics{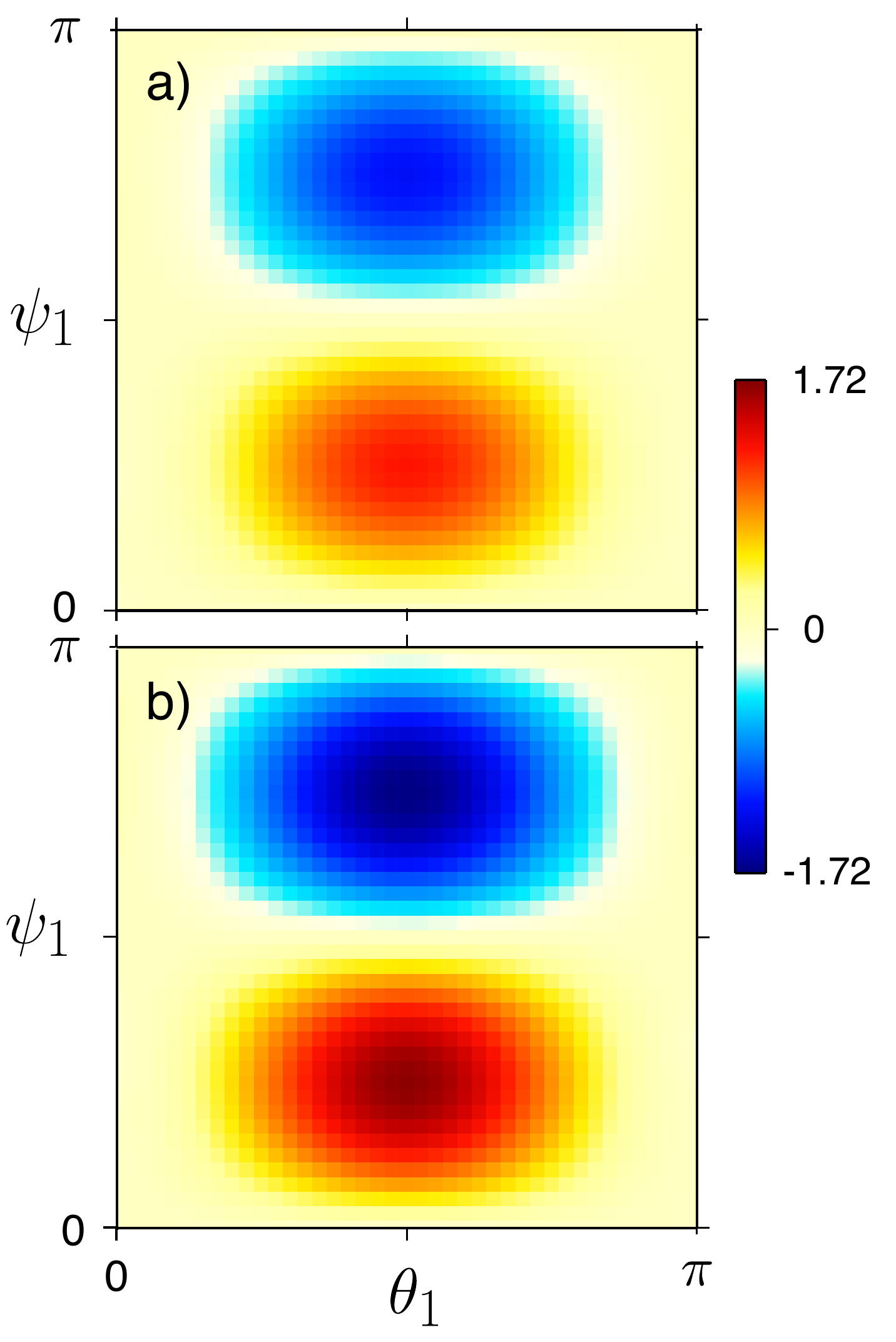}}
\end{center}
\caption{\label{fig_psdotso}
Instantaneous crystallographic spin $\dot\psi^{[1]}(\theta^{[1]}, \psi^{[1]})$ for the slip system (010)[100], predicted by
the SO model with $\tau^{[1]}/\tau^{[2]} = \tau^{[2]}/\tau^{[3]} = 1$ for an initially
isotropic olivine aggregate deformed in uniaxial compression. (a)
$|\dot{\epsilon_3}|t= 0$;  (b) and $|\dot{\epsilon_3}|t= 0.4$.
Color scale is in units of the axial shortening rate $\dot{\epsilon}_3 < 0$). The
Eulerian angles $\theta^{[1]}$ and $\psi^{[1]}$ are defined as in
Fig. \ref{fig_eulerangles}b.
}
\end{figure}

To go further, we first note the
obvious difficulty that the space of possible background textures
$f$ is infinite.  To make progress, therefore, we need to restrict 
and parameterize this space in some way. Our choice is to consider
the space of all textures produced by uniform triaxial straining of an initially isotropic aggregate,
which can be parameterized by the axial ratios $r_{12}$ and $r_{23}$
of the associated FSE. Accordingly, the functional dependence we need to determine
becomes 
\be
\dot\gamma^{[s]}=\dot\gamma^{[s]}\left(\ve g^{[s]},\mathbf{E}, r_{12}, r_{23}, p_{12}, p_{23}, m\right).
\label{gdot2}
\ee
This still seems impossibly complex, so we now call the SO model to
our aid. Consider the case of uniaxial compression along the 
$x_3$-axis at a rate $\dot{\epsilon}$, for which the nonzero components of the strain rate tensor are
$E_{33}= - \dot\epsilon$, $E_{11} = E_{22} = \dot\epsilon/2$. 
The shear rate $\dot{\gamma}^{[1]}$
and the ODF $f$ are then independent of the Eulerian angle $\phi^{[1]}$ by symmetry.
Fig. \ref{fig_psdotso} shows the spin $\dot\psi^{[1]}(\theta^{[1]}, \psi^{[1]})$
for the slip system  (010)[100] ($s=1$) predicted by the
SO model  with $p_{12} = p_{23} = 0$ ($\tau^{[1]} = \tau^{[2]} = \tau^{[3]}$) at two
different equivalent strains
$r_0\equiv |\dot{\epsilon_3}| t=0$ (Fig. \ref{fig_psdotso}a) and
$r_0=0.4$ (Fig. \ref{fig_psdotso}b).
Remarkably, the images of Figs. \ref{fig_psdotso}a
and Figs. \ref{fig_psdotso}b appear to be the
same function with different amplitudes. A more detailed
investigation shows that this impression is correct, and that  
the function in question is
$F = b \sin 2\psi^{[1]} \sin^2\theta^{[1]}$, where $b$ is
an unknown amplitude.  Least-squares fitting
of this expression to the numerical predictions yields $b = 1.25$ 
for  fig.~\ref{fig_psdotso}a and $b = 1.71$ for fig.~\ref{fig_psdotso}b,
with a nearly perfect fit (variance reduction = 99.9\%) in both cases.

Next, we note that the function
$\sin 2\psi^{[1]} \sin^2\theta^{[1]}$ can be written as
\be
\sin 2\psi^{[1]} \sin^2\theta^{[1]} = - \frac{2\sqrt{3}}{3} T_2^{\prime\prime 20} = -2 (S_{11}^{[1]} + S_{22}^{[1]}),
\label{ghsform}
\ee
where $T_2^{\prime\prime 20}$ is a generalized spherical harmonic (GSH), defined as 
$T_l^{\prime\prime mn} = 2^{-1/2} i^{1+m-n}
\left(
T_l^{mn} - T_l^{-m-n}
\right)$
(Bunge 1982, eqn. 14.37), where $i = \sqrt{-1}$. 
This result has two surprising and far-reaching implications. 
First, the angular dependence of the spin ($\propto \sin 2\psi^{[1]} \sin^2\theta^{[1]}$ 
in this case) remains the same regardless of the 
strength of the background texture; it is only the amplitude of the function
that depends on the texture. Second, it suggests that
the angular dependence of the spin is always a GSH
of degree $l=2$, without any contribution from
higher-degree harmonics.   Noting further that the 
shear rate $\dot{\gamma}^{[s]}$ must depend linearly on the imposed macroscopic strain rate $\mathbf{E}$, we are led to propose the following expression for $\dot{\gamma}^{[s]}$:
\be\dot{\gamma}^{[s]}=-\dot{\psi}^{[s]}=A_{ijkl}^{[s]}(r_{12},r_{23},p_{12}, p_{23}, m) S^{[s]}_{ij}E_{kl},
\label{gammadotGeneral}
\ee
where $\mathbf{A}$ is a fourth-order `spin tensor'.
The superscripts
$[s]$ denote the index of the slip system ($s = 1$, 2 or 3).
In the next section we determine how the spin tensor
$A_{ijkl}$ depends on its five arguments.

\section{Parameterisation of the spin tensor $\mathbf{A}$}

\subsection{SO model calculations}

We now use the SO model as a benchmark to determine the 
tensor components $A_{ijkl}(r_{12},r_{23},p_{12}, p_{23}, m)$.
The procedure comprises two steps: 
(1) generation of the 
background texture and (2) calculation of the instantaneous
spin induced by applying a given rate of strain
to the background texture. 
Thus in step (1),  we first select the number of crystals $N$
in the model aggregate (= 2000 in all cases) and the values of the slip-system parameters
$p_{12}$,  $p_{23}$, and 
$m$ (= 3.5 in all cases). We also choose the components
of the strain rate tensor  $\boldsymbol{\cal{E}}$ that
generates the background texture. We work in the reference
frame of the FSE, which means that 
\be 
\boldsymbol{\cal{E}} =
\left(
\begin{array}{ccc}
{\cal E}_{11}  &  0 &  0\\
0  & {\cal E}_{22}  & 0 \\
0  &  0 & {\cal E}_{33}\\
\end{array}
\right)
\ee
The SO model is then run
starting from an isotropic initial condition until target values
of the FSE axial ratios $r_{12}$ and $r_{23}$ are reached. 
In step 2, we apply an instantaneous 
strain rate tensor 
\be
\mathbf{E} =
\left(
\begin{array}{ccc}
E_{11}  &  E_{12} &  E_{13}\\
E_{12}  & E_{22}  & E_{23} \\
E_{13}  &  E_{23} & E_{33}\\
\end{array}
\right)
\ee
to the background texture. Note that $\mathbf{E}$
need not be the same as $\boldsymbol{\cal{E}}$,
which allows us to obtain results for arbitrary orientations of the principal axes
of $\mathbf{E}$ relative to those of $\boldsymbol{\cal{E}}$.

The final result of the procedure described above is a set of 
slip rates $\dot{\gamma}_n^{[s]}$
on each of the three slip systems ($s=1,2$ or 3 in Table 1) and
for each of the $n$ phases ($n = 1, 2, ...., N$). 
The calculated values of  $\dot{\gamma}_n^{[s]}$
are then substituted into (\ref{omega}) to obtain
the `partial' spins $\omega^{[s]}_i$ due to the actions of the individual
slip systems, and which are related to the total spin $\omega_i$ by
\be
\omega_i = \sum_{s=1}^3\omega^{[s]}_i.
\ee
Finally, by substituting the partial spins $\omega^{[s]}_i$ into (\ref{gdotdef})
and expressing the results in terms of the slip system-specific Eulerian angles
$(\phi^{[s]},\theta^{[s]}, \psi^{[s]})\equiv\ve g^{[s]}$, we obtain the
rotation rates $\dot{\psi}_n^{[s]}$ for all grains $n$ and slip systems $s$.

\subsection{Analytical parameterisation}

At this point, we have a large library of numerical solutions, but little
idea of what they imply about the structure of the function 
$A^{[s]}_{ijkl}(r_{12},r_{23},p_{12},p_{23}, m)$. 
As a first simplification, we assume that
\be
A^{[s]}_{ijkl}=H^{[s]}(p_{12},p_{23}, m) Q_{ijkl}(r_{12},r_{23}).
\label{bigacompact}
\ee
The first factor $H^{[s]}$ in (\ref{bigacompact}) describes how the 
activities of the three slip systems 
depend on the CRSS ratios at the initial
instant $(r_{12} = r_{23} = 0)$ of the deformation, whilst the factor $Q_{ijkl}(r_{12},r_{23})$ describes how the activities of slip systems with equal strengths ($p_{12} = p_{23} = 0$)
vary as a function of strain for arbitrary deformations. 

Consider first the factor $Q_{ijkl}(r_{12},r_{23})$. 
Since $E_{ij}$ and $S_{ij}^{[s]}$ are both symmetric and traceless, there are at most 25 independent
products of them, or equivalently 25 independent $Q_{ijkl}$. 
However, we have found that $Q_{ijkl}(r_{12},r_{23})$ obeys 
surprising symmetries that reduce the number of its
independent non-zero components to just two. We began by fixing $H^{[s]}(0,0,3.5)=1$ and
performing a least-squares fit of the model (\ref{gammadotGeneral}) to the spin predicted by the SO model, for random values of $r_{12}$ and $r_{23}$, and random instantaneous strain-rate tensors (SRTs). 
This allowed us to discover numerically that eighteen of the coefficients $Q_{ijkl}$ were identically zero.
We also found at this stage that $Q_{1122}\approx Q_{2211}$. 
These two numerical results implied that the tensor $Q_{ijkl}$ exhibits \textit{major} symmetry, i.e. $Q_{ijkl}=Q_{klij}$. This leaves only six independent, non-zero components of $Q_{ijkl}$, namely $Q_{1111}$, $Q_{1122}$, $Q_{2222}$, $Q_{1212}$, $Q_{1313}$ and $Q_{2323}$.

Relationships among the six remaining non-zero
$Q_{ijkl}$ arise from the fact that the labelling of the coordinate
axes is arbitrary. We can have a cyclic permutation of the coordinate axes from $(1,2,3)$ to $(2,3,1)$ or $(3,1,2)$, or a non-cyclic permutation from $(1,2,3)$ to $(1,3,2)$, $(2,1,3)$ or $(3,2,1)$. The spin $\dot{\psi}^{[s]}$
has to be invariant under a relabeling of the coordinate
axes. Equating the expressions for $\dot{\psi}^{[s]}$ in the original
and the transformed coordinate systems allows us to derive rigorous transformation rules (in the reference frame of the FSE). Setting $B=Q_{1111}$ and $C=Q_{1212}$, we find that
\bsub
\begin{eqnarray}
Q_{1212}\left(r_{12},r_{23}\right)
&=&C\left(r_{12},r_{23}\right),\\
Q_{1313}\left(r_{12},r_{23}\right)
&=&C\left(r_{31},r_{23}\right),\\
Q_{2323}\left(r_{12},r_{23}\right)
&=&C\left(r_{23},r_{12}\right),\\
Q_{1111}\left(r_{12},r_{23}\right)
&=&B\left(r_{12},r_{23}\right),\\
Q_{2222}\left(r_{12},r_{23}\right)
&=&B\left(r_{12},r_{31}\right),\\
Q_{1122}\left(r_{12},r_{23}\right)&=&\tfrac{1}{2}\left[B\left(r_{23},r_{12}\right)-B\left(r_{12},r_{23}\right)-B\left(r_{12},r_{31}\right)\right].
\end{eqnarray}
\label{symrules}
\esub
These symmetries that we have
uncovered reduce the number of independent 
coefficients $Q_{ijkl}$ to just two, which we take to be $Q_{1111}$ and $Q_{1212}$. 

Now consider the factor $H^{[s]}$ in (\ref{bigacompact}). 
We have discovered numerically that $H^{[2]}$ and $H^{[3]}$ can be obtained from
$H^{[1]}$ by simple variable transformations:
\be
H^{[2]}(p_{12},p_{23})= H^{[1]}(-p_{12},-p_{31}),
\quad\quad
H^{[3]}(p_{12},p_{23}) = H^{[1]}(-p_{23},-p_{12}).
\label{hrot}
\ee
Combining this with the previous results for $Q_{ijkl}$, we obtain the following 
model for the crystallographic spin, which is valid on each slip system $s$:
$$
\dot{\psi}^{[s]} = 
\frac{1}{2}H^{[s]}(p_{12},p_{23})\Big{\{}
$$
$$
B\left(r_{12},r_{23}\right)
\big{[}\left(-4E_{11}+E_{22}\right)S^{[s]}_{11}+\left(E_{11}+2E_{22}\right)S^{[s]}_{22}
\big{]}
$$
$$
-B\left(r_{23},r_{12}\right)
\big{[}\left(4E_{11}+5E_{22}\right)S^{[s]}_{11}+\left(5E_{11}+4E_{22}\right)S^{[s]}_{22}
\big{]}
$$
$$
+B\left(r_{12},r_{31}\right)
\big{[}\left(2E_{11}+E_{22}\right)S^{[s]}_{11}+\left(E_{11}-4E_{22}\right)S^{[s]}_{22}
\big{]}
$$
\be
\label{PSIDOT1}
\quad\;-8\left[C\left(r_{12},r_{23}\right)E_{12}S^{[s]}_{12}
+C\left(r_{23},r_{12}\right)E_{23}S^{[s]}_{23}
+C\left(r_{31,}r_{23}\right)E_{31}S^{[s]}_{31}\right]
\Big{\}}.
\ee
Note that the coefficient $C$ multiplies the off-diagonal components
of $\mathbf{E}$, and is therefore not needed for coaxial deformations where the principal axes of the SRT and the FSE are aligned. 

\subsection{Numerical determination of the parameterisation coefficients}

\begin{figure}\begin{center}\scalebox{0.75}{\includegraphics{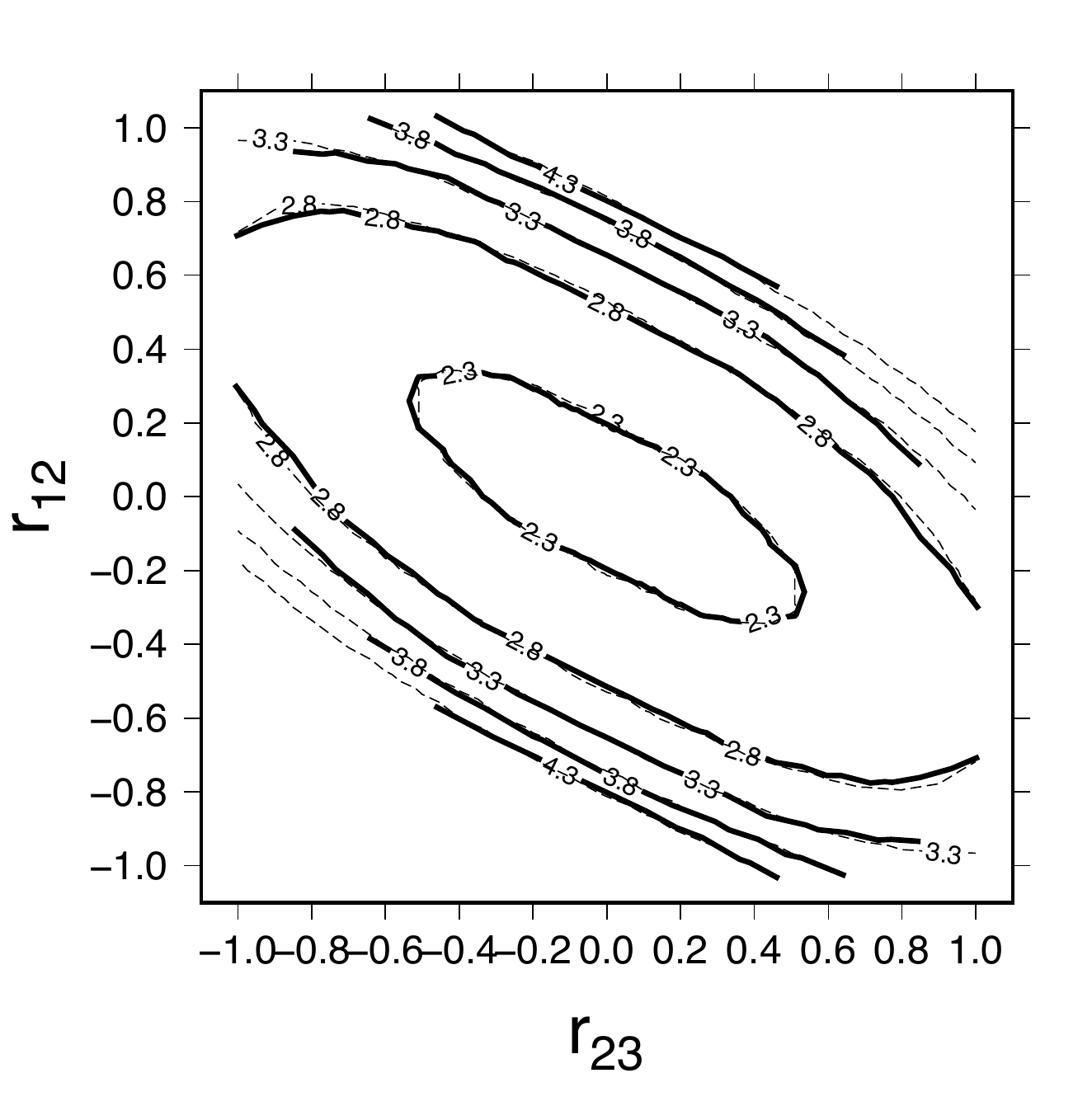}}\end{center}
\caption{\label{fig_A1111} Spin amplitude $B$ as
a function of deformation when the strengths of the dominant slip systems
are all equal, shown as a function
of the axial ratios of the finite strain ellipsoid. 
Solid contours show the amplitude $B$ that best fits
the predictions of the SO model, and the dashed contours
show the fitting function (\ref{zmn}).}\end{figure}

\begin{figure}\begin{center}\scalebox{0.75}{\includegraphics{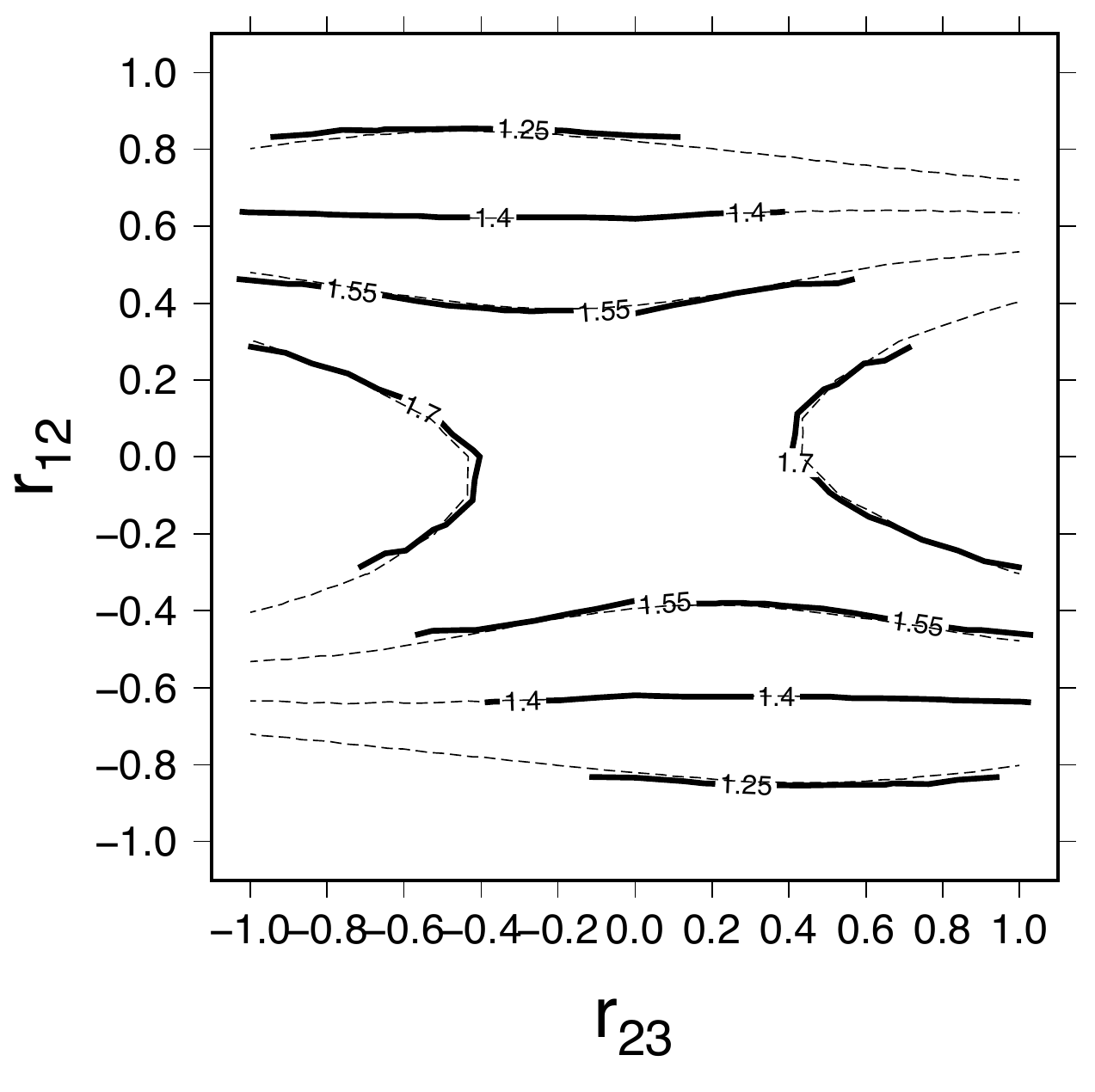}}\end{center}
\caption{\label{fig_A1212} 
Same as fig.~\ref{fig_A1111}, but for the spin amplitude $C$.}
\end{figure}

The symmetries outlined above indicate that we only require analytical expressions for the three functions $B$, $C$ and $H^{[1]}$. Full details of the expressions obtained and the methods used are given in Appendix B.
Briefly, we first obtain models for $B$ and $C$, by setting equal slip-system strengths 
($p_{12} = p_{23} = 0$) and fixing $H^{[1]}(0,0)=1$ in (\ref{PSIDOT1}). We then capture $B$ and $C$ data by a least-squares fit of the model (\ref{PSIDOT1}) to the spin predicted by the SO model, for sampled values of $r_{12}$ and $r_{23}$, and random instantaneous SRTs. In each case, the variance reduction of the fit $R\geq 99.7\%$.  Simple polynomials in $r_{12}$ and $r_{23}$ are fitted to the $B$ and $C$ data (using least squares) to obtain the analytical expressions (\ref{zmn}). The RMS errors of the fits are $0.039$ and $0.0070$, respectively. Figs \ref{fig_A1111} and \ref{fig_A1212} display contour plots of the models (\ref{zmn}) against the $B$ and $C$ data, respectively.

To obtain an analytical expression for $H^{[1]}$, we capture 
$H^{[1]}$ data over the entire range $[-\ln4,\ln4] =[-1.386,1.386]$  of $p_{12}$ and $p_{23}$ for olivine.
We fit (\ref{PSIDOT1}) to 81 instantaneous ($t = 0$) numerical solutions of the SO model for uniaxial compression, with equally spaced points in the $(p_{12},p_{23})$-plane with $p_{12}$ and $p_{23}$ in the range [-1.386,1.386].
Simple polynomials in $p_{12}$ and $p_{23}$ are fitted to the $H^{[1]}$ data using least squares,
leading to the analytical expression (\ref{A1p}). The RMS error of the fit is 0.0068. 

Finally, we test the assumption (\ref{bigacompact})
that $A_{ijkl}$ can be written as a product of a scalar $H^{[s]}$ (that depends
on $p_{12}$ and $p_{23}$) and a tensor $Q_{ijkl}$ (that depends on $r_{12}$ and $r_{23}$). We substituted the analytical expressions for $B$, $C$ and $H^{[1]}$ (see (\ref{zmn}) and (\ref{A1p}), respectively) into the full model for the spin on each slip system (\ref{PSIDOT1}). We then fitted these models to the spin predicted by the SO model for random background textures (formed from various $r_{12}$, $r_{23}$, $p_{12}$ and $p_{23}$ values) and random instantaneous SRTs. Remarkably, in each case, the variance reduction $R> 99.1\%$ and in most cases $R>99.7\%$. 

\section{Evolution of CPO during progressive deformation}

The results in the previous sections imply that the ANalytical PARameterization (`ANPAR' model) 
provides an accurate and efficient substitute for the much more computationally
expensive SO model. We now demonstrate this in more detail
by comparing the textures predicted by the two models for
olivine polycrystals subjected to various kinds of finite deformation. In the first three test cases (for uniform deformation) the strain increment used $\left(\Delta r_0 =0.025\right)$ is the same for both the SO and ANPAR models. The different components of our method for ANPAR CPO calculation are summarized in Algorithm 1.

\line(1,0){450}\\
\textbf{Algorithm 1} ANPAR CPO calculations for Olivine
\\\line(1,0){450}\\
(1) Set $N$ \hfill (Number of grains ($n=1,...,N$))\\
(2) Set initial isotropic texture $\ve g_n[0]=(\phi_n[0], \theta_n[0],\psi_n[0])$.\\
(3) Set $p_{12},p_{23}$\hfill (CRSS ratios)\\
(4) Set $r_{12}[0]=r_{23}[0]=0$. \hfill (FSE initially a sphere)\\
(5) For $k=1,...,K$ \textbf{do}\\
\indent\indent(i) Calculate $\mathbf E[k]$ and $\mathbf \Omega[k]$ \hfill (SRT and macroscopic rotation rate)\\
\indent\indent(ii) Transform SRT into reference frame of FSE\\
\indent\indent(iii) Calculate $\mathbf D[k]$ \hfill (velocity gradient tensor, using (\ref{bigdij}))\\
\indent\indent(iv) Set $r_0[k]$ \hfill (strain increment)\\
\indent\indent(v) Calculate $r_{12}[k]$, $r_{23}[k]$ and $t_k$ \hfill (FSE parameters and length of\\
\indent\indent time-step, using (\ref{r0}))\\
\indent\indent(vi) Calculate slip-rates $\dot{\gamma}_n^{[s]}[k]$ \hfill (using (\ref{psidoteqgammadot}) and (\ref{PSIDOT1}))\\
\indent\indent(vii) Calculate rotation rates $\dot{\ve g}_n[k]=\left(\dot{\phi}_n[k],\dot{\theta}_n[k],\dot{\psi}_n[k]\right)$ \hfill (using (\ref{gdotdef}))\\
\indent\indent(viii) Update texture $\ve g_n[k]=\left(\phi_n[k],\theta_n[k],\psi_n[k]\right)$ \hfill(integrating\\
\indent\indent forward in time)\\
     \indent\textbf{end do}\\
(6) Plot texture $\ve g_n[K]=\left(\phi_n[K],\theta_n[K],\psi_n[K]\right)$ \hfill (using MTEX)
\line(1,0){450}

\subsection{Irrotational deformations}

Our first test case is uniform deformation by
uniaxial compression to a strain $r_{23} = 0.9$, $r_{12} = 0$, with CRSS ratios $\tau^{[1]}/\tau^{[2]} = 0.5$ and $\tau^{[2]}/\tau^{[3]} = 0.667$.
Fig. \ref{fig_uniax} shows the (100), (010)
and (001) pole figures predicted for this case by the
SO model (Fig. \ref{fig_uniax}a) and the ANPAR model
(Fig. \ref{fig_uniax}b). The two sets of pole figures 
are practically indistinguishable (variance reduction $R>99.9\%$).
However, the ANPAR model is a remarkable $1.75\times10^4$ times quicker than the SO model (0.0344 s for ANPAR vs. 603 s for SO). 

To quantify the agreement in another way, we used the orientations to calculate the polycrystalline
elasticity of the aggregate for the SO and ANPAR models using the MSAT
software (Walker and Wookey, 2012). An element-by-element comparison of
the two Voigt-Reuss-Hill average elastic stiffness tensors gives a maximum
absolute difference of 0.19 GPa between SO and ANPAR, which is not significant for geophysical
purposes.

As a second test, Fig. \ref{fig_pure} shows the predicted pole figures for
uniform deformation by pure shear in the $x_1$-$x_3$ plane
to a strain $r_{12}= r_{23} = 0.563$, again with CRSS ratios $\tau^{[1]}/\tau^{[2]} = 0.5$ and $\tau^{[2]}/\tau^{[3]} = 0.667$. 
Again, the two sets of pole figures and the predicted elasticity are
nearly identical (variance reduction $99.3\%$,  maximum absolute difference in the predicted
elasticity 0.16 GPa). In this case, the speed of 
the ANPAR model is 3.1$\times 10^4$ greater than that of
the SO model (0.0348 s for ANPAR vs. 1090 s for SO). 

\subsection{Rotational deformations}

Rotational deformations are those in which the axes of the 
FSE do not remain aligned with
the principal axes of the SRT as the deformation progresses.

As an example, consider the case of simple shear, for which
the major axis of the FSE is initially aligned with the SRT
but then rotates progressively away from it towards the shear plane.  
As a result, both functions $B$ and $C$ in (\ref{PSIDOT1}) come into play.

 Let $\dot{\epsilon}_1$ be the maximum rate of extension along the $x_1$ axis. The elongation of the FSE at time $t$ can then be described by the axial ratio ${\cal R}=\exp(r_{12})=\exp(2\dot{\epsilon}_1t)$. If we denote
$\chi(t)$ the angle between the two frames, then $\chi(0) = 0$
and $\lim_{t\rightarrow\infty}\chi(t) = - \pi/4$. Using the standard tensor
transformation rule, we obtain a velocity gradient tensor of the form $\mathbf{D}=\mathbf{E}+\mathbf{W}$, where
\be 
\mathbf{E} =
\dot{\epsilon}_1\left(
\begin{array}{ccc}
\cos2\chi  & \sin2\chi  &  0\\
\sin2\chi  & -\cos2\chi  & 0 \\
0  &  0 & 0\\
\end{array}
\right),
\quad\quad
\mathbf{W} =
\dot{\epsilon}_1\left(
\begin{array}{ccc}
0  & -1 &  0\\
1 & 0  & 0 \\
0  &  0 & 0\\
\end{array}
\right)
\ee 
are the respective strain-rate and rotation-rate, tensors and\\ $\chi =-\tfrac{1}{2}\tan^{-1}\left(\dot{\epsilon}_1t\right)$. We used CRSS ratios $\tau^{[1]}/\tau^{[2]} = 0.5$ and $\tau^{[2]}/\tau^{[3]} = 0.667$. We updated the velocity gradient tensor at each time step to remain in the frame of reference of the FSE.

Fig. \ref{fig_sshear} shows the pole figures predicted by our theory
together with those predicted by the SO model for $\dot{\epsilon}_1=1$ and $r_0=0.5$. Yet again, the 
two sets of pole figures are nearly indistinguishable, with a variance
reduction $R=99.5\%$ and a maximum difference in the predicted elasticity of
0.88 GPa. The speed of 
the ANPAR model is 7.6$\times 10^4$ greater than that of
the SO model (0.273 s for ANPAR vs. 2062 s for SO). 

\subsection{Non-Newtonian corner-flow model for a spreading ridge}

\begin{figure}\begin{center}\scalebox{0.65}{\includegraphics{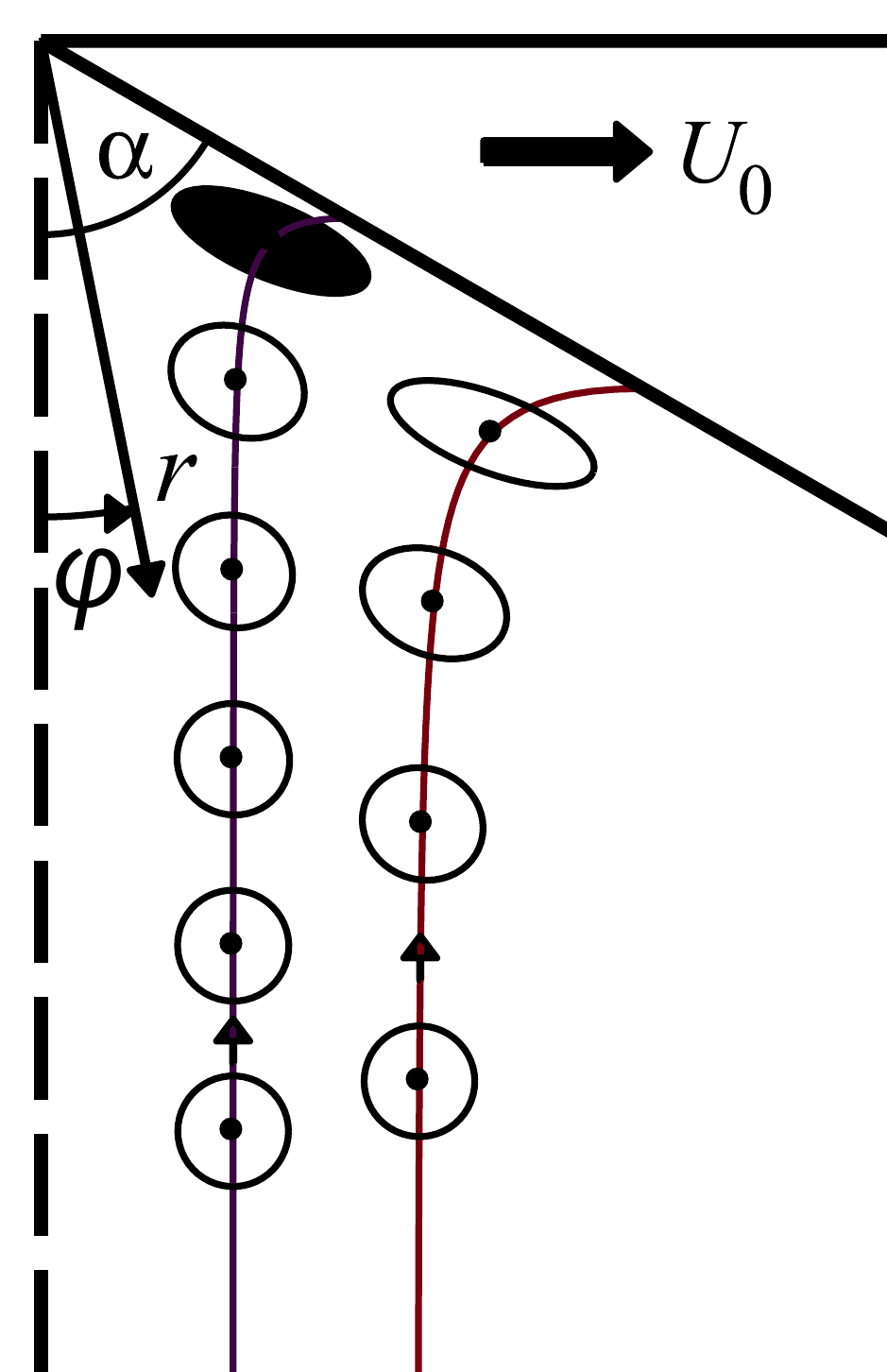}}\end{center}
\caption{\label{CF_geom}Geometry and boundary conditions for the corner flow model of a ridge crest (right half only). The half spreading rate is $U_0$ and the asthenospheric wedge has angular dimension $\alpha =60^{\circ}$. Solid lines with arrows show typical streamlines for a power law rheology with index $n=3$. The FSE's are plotted at different points along the two streamlines. The black ellipse relates to the point in the flow at which the texture is approximated (see Fig. \ref{fig_CF}).
}\end{figure}

\begin{figure}\begin{center}{\includegraphics[scale=0.7]{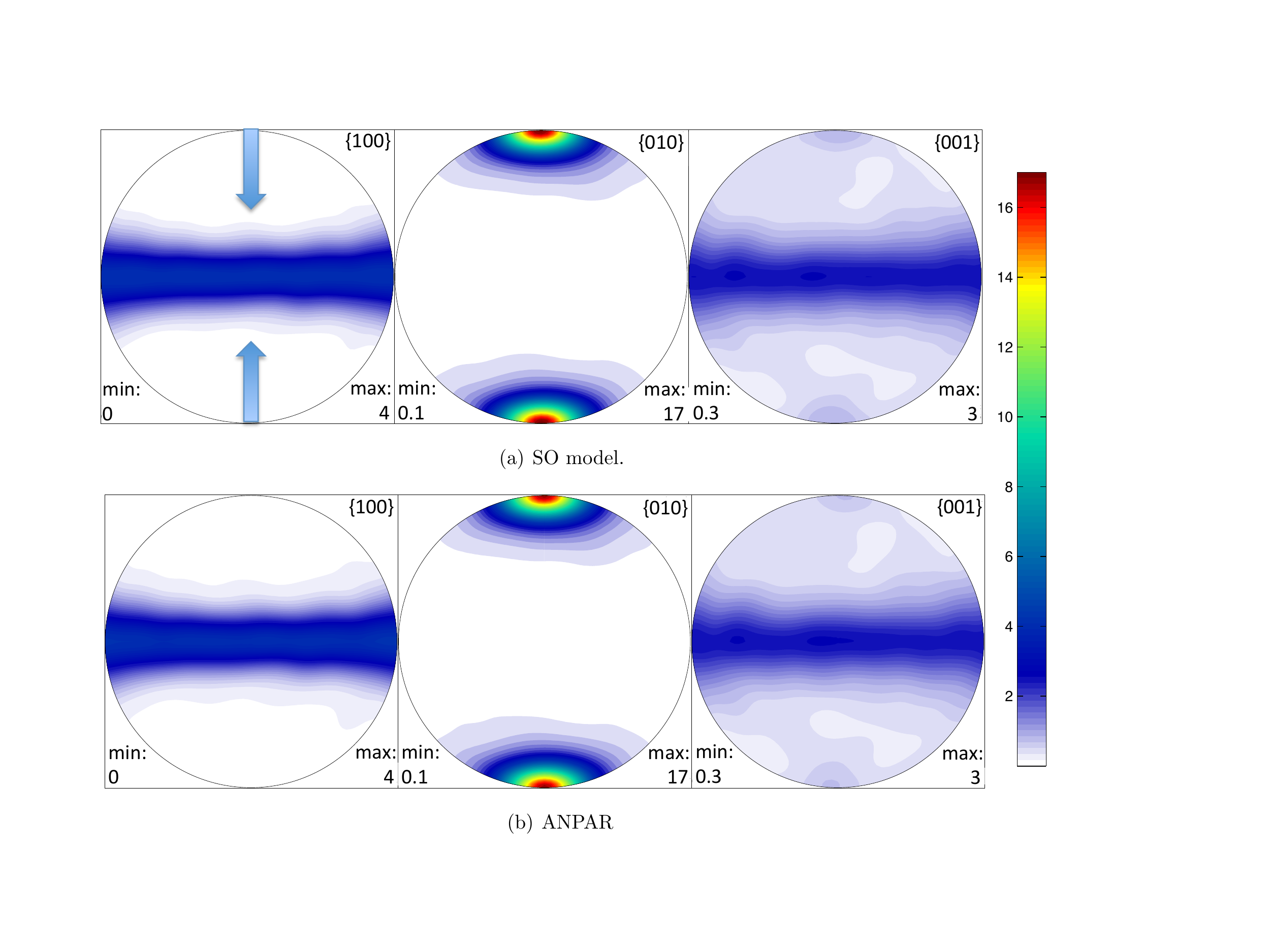}}\end{center}\caption{Pole figures (equal-area projections) for an olivine polycrystal deformed
by uniaxial compression to a strain $r_{23} = 0.9$, $r_{12} = 0$.
The arrows indicate the compression ($x_3$-) axis, which extends from the bottom to the top of each figure. The predictions of (a) the SO model and (b) the analytical model (ANPAR) are shown
for critical resolved shear stress ratios $\tau^{[1]}/\tau^{[2]} = 0.5$ and $\tau^{[2]}/\tau^{[3]} = 0.667$. Figure generated using MTEX (Bachmann et al., 2010).}\label{fig_uniax}\end{figure}

\begin{figure}\begin{center}
{\includegraphics[scale=0.7]{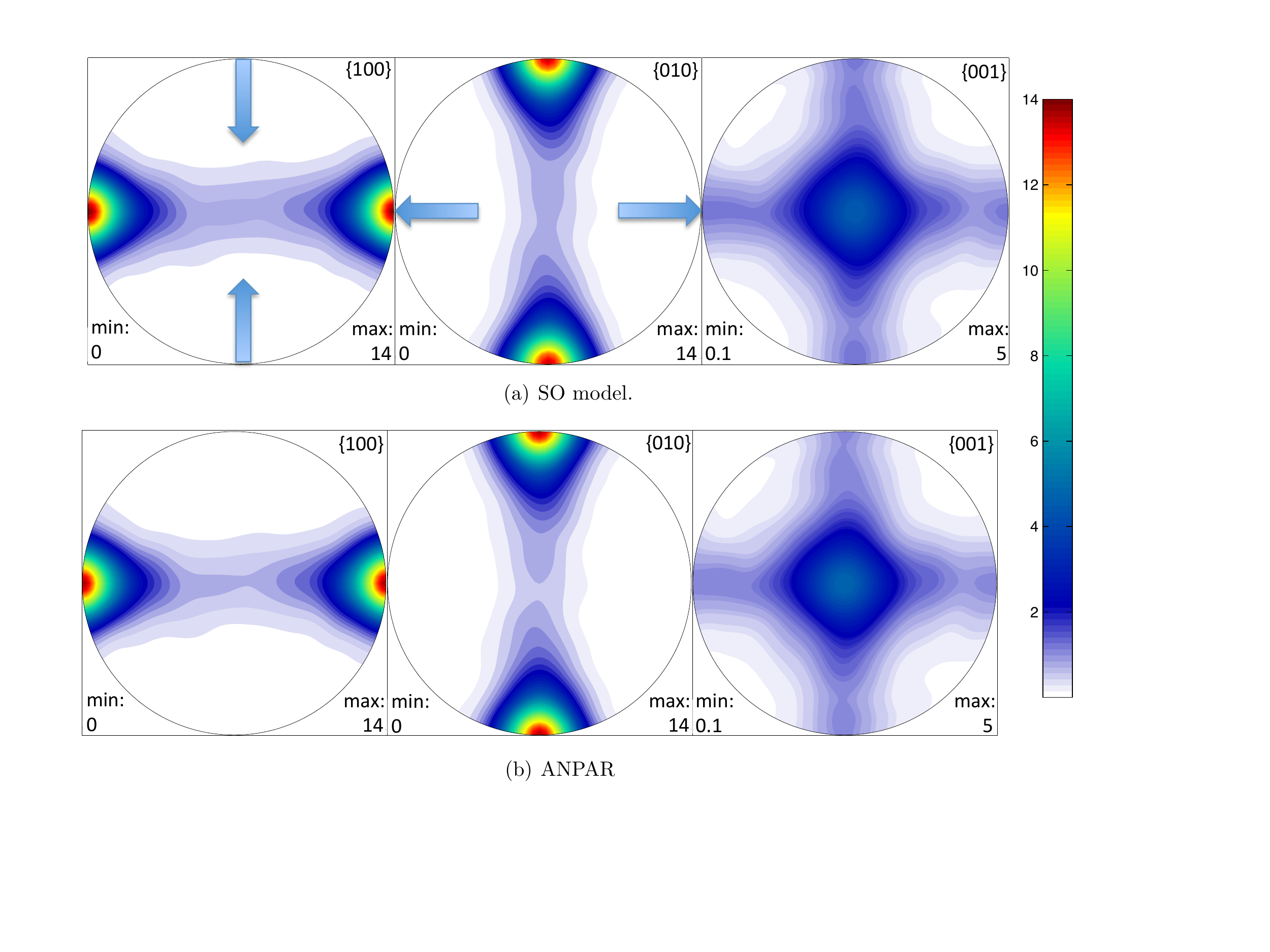}}
\end{center}
\caption{Same as Fig. \ref{fig_uniax}, but for deformation by pure shear to 
$r_{12} = r_{23} = 0.563$.
The axes of maximum extension ($x_1$) and compression ($x_3$) are indicated by the arrows.}\label{fig_pure}\end{figure}

\begin{figure}\begin{center}
{\includegraphics[scale=0.7]{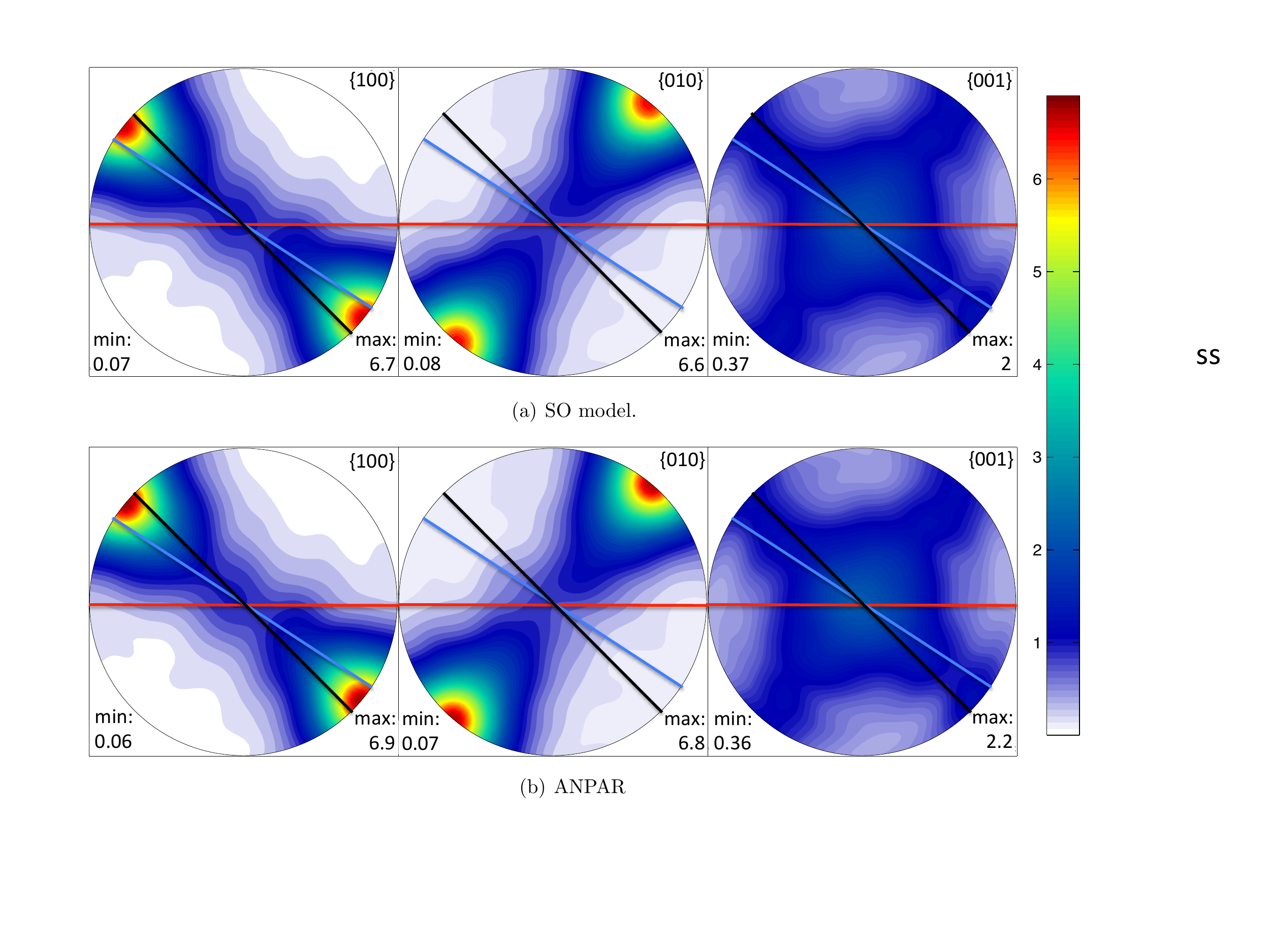}}
\end{center}
\caption{Same as Fig. \ref{fig_uniax}, but for deformation by simple shear to 
$r_{12} = r_{23} = 0.433$ ($r_0 = 0.5$), with CRSS ratios $\tau^{[1]}/\tau^{[2]} = 0.5$ and $\tau^{[2]}/\tau^{[3]} = 0.667$. The black, red and blue lines indicate the axis of maximum instantaneous extension ($x_1$), the shear plane and the long axis of the FSE, respectively.}\label{fig_sshear}\end{figure}

Our final example is a more complex
and non-uniform geophysical flow, namely the flow in the mantle beneath an ocean ridge.
This flow can be simply modeled using the `corner flow' similarity solution of the Stokes equation in polar coordinates $(r,\varphi)$ (Batchelor, 1967). Figure \ref{CF_geom} shows the geometry and boundary conditions appropriate for a ridge crest (Lachenbruch $\&$
Nathenson, 1974). Flow in the asthenosphere $0<\varphi<\alpha$ is driven by the horizontal motion of wedge-shaped surface plates at velocity $U_0$. The solid lines with arrows show typical streamlines of the flow for an asthenosphere with a power law rheology with power law index $n=3$
(Tovish et al., 1978).
The two streamlines are for $\varphi_0=10^\circ$ and $20^\circ$, and we use $\alpha=60^\circ$ throughout this paper.

The steady incompressible Stokes equations and the boundary conditions in Fig. \ref{CF_geom} can be satisfied if the stream function $\Psi$ has the self-similar form
\be
\Psi=U_0 r F(\varphi),
\label{StreamFn}
\ee
which is valid for both Newtonian $(n=1)$ and non-Newtonian ($n\neq 1$) fluids. 
Here we use $n=3$, corresponding to a rheology that is close to that of olivine
at high stresses ($n\approx 3.5$; Bai et al. 1991).
The function $F(\varphi)$ for $n=3$ is of the form
\be
F(\varphi)=A\sin\varphi+Ch(\varphi,D),
\ee
where
\be
h(\varphi,D)=27\cos\left[\frac{\sqrt{5}}{3}(\varphi+D)\right]-\cos\left[\sqrt{5}(\varphi+D)\right]
\ee
The constants $A,C$ and $D$ are chosen to satisfy the boundary conditions, yielding
\begin{eqnarray}
D&=&\frac{3\pi}{2\sqrt{5}},\\
C&=&-\left[h(\alpha,D)\cos\alpha-h_\varphi(\alpha,D)\sin\alpha\right]^{-1},\\
A&=&-C\left[h(\alpha,D)\sin\alpha+h_\varphi(\alpha,D)\cos\alpha\right],
\end{eqnarray}
where $h_\varphi=\mathrm{d} h/\mathrm{d}{\varphi}$.
The maxium strain rate $\dot{\epsilon}$ is
\be
\dot{\epsilon}=U_0\frac{|F^{\prime\prime}+F|}{2r}
\ee
and the local rotation rate (= one-half the vorticity) is 
\be\label{OmCF}
\Omega=-U_0\frac{F^{\prime\prime}+F}{2r}.
\ee

To proceed we require knowledge of the FSE as we progress along a streamline. We obtain the axial ratio ${\cal R}=\exp(r_{12})$ of the FSE and the orientation $\chi$ of the FSE by solving the following evolution equations (Kellogg and Turcotte 1990; Ribe 1992):
\begin{eqnarray}
\dot{\cal R}&=&2{\cal R}\left(E_{11}\cos2\chi+E_{12}\sin2\chi\right),
\\\dot{\chi}&=&\Omega+\frac{1+{\cal R}^2}{1-{\cal R}^2}\left(E_{11}\sin2\chi-E_{12}\cos2\chi\right).
\label{EvEqn1}\end{eqnarray}

The above equations can be simplified by transforming the Cartesian strain
rate components $E_{ij}$ to polar coordinates,  and then expressing the time derivatives in terms of a $\varphi$-derivative (McKenzie 1979, eqn. 6):
\be
\frac{D}{Dt}=-\frac{U_0 F} {r}\frac{d}{d\varphi}.
\ee
This leads to the following simplified form for the evolution equations: 
 \begin{eqnarray}
\frac{d\cal R}{d\varphi}&=&-{\cal R}\;\frac{F^{\prime\prime}+F}{F}\sin2\left(\chi-\varphi\right),\\
\frac{d\chi}{d\varphi}&=&\frac{F^{\prime\prime}+F}{2F}\left[1+\frac{1+{\cal R}^2}{1-{\cal R}^2}\cos2\left(\chi-\varphi\right)\right]
\end{eqnarray}
which must be solved subject to the following initial conditions at $\varphi = \varphi_0$:
\begin{eqnarray}
{\cal R}\left(\varphi_0\right)=1,\\
\chi\left(\varphi_0\right)=\varphi_0+ \frac{\pi}{4}.
\end{eqnarray}
These evolution equations were solved using a fourth order Runge-Kutta method. In Fig (\ref{CF_geom}), the FSE is plotted at different points along two different streamlines.

To calculate CPO, we have to transform back into Cartesian coordinates. In doing this, we obtain a velocity gradient tensor of the form $\mathbf{D}=\mathbf{E}+\mathbf{W}$, where
\be 
\mathbf{E} =
\dot{\epsilon}\left(
\begin{array}{ccc}
-\sin2\varphi & \cos2\varphi  &  0\\
\cos2\varphi  & \sin2\varphi & 0 \\
0  &  0 & 0\\
\end{array}
\right),
\quad\quad
\mathbf{W} =
\dot{\epsilon}\left(
\begin{array}{ccc}
0  & 1 &  0\\
-1 & 0  & 0 \\
0  &  0 & 0\\
\end{array}
\right)
\ee 

We then transform the SRT into the reference frame of the FSE. This gives 
\be 
\mathbf{E} =
\dot{\epsilon}\left(
\begin{array}{ccc}
\sin 2 (\chi-\varphi)  & \cos 2 (\chi-\varphi)  &  0\\
\cos 2 (\chi-\varphi)  & -\sin 2 (\chi-\varphi) & 0 \\
0  &  0 & 0\\
\end{array}
\right)
\ee

\begin{figure}\begin{center}
{\includegraphics[scale=0.7]{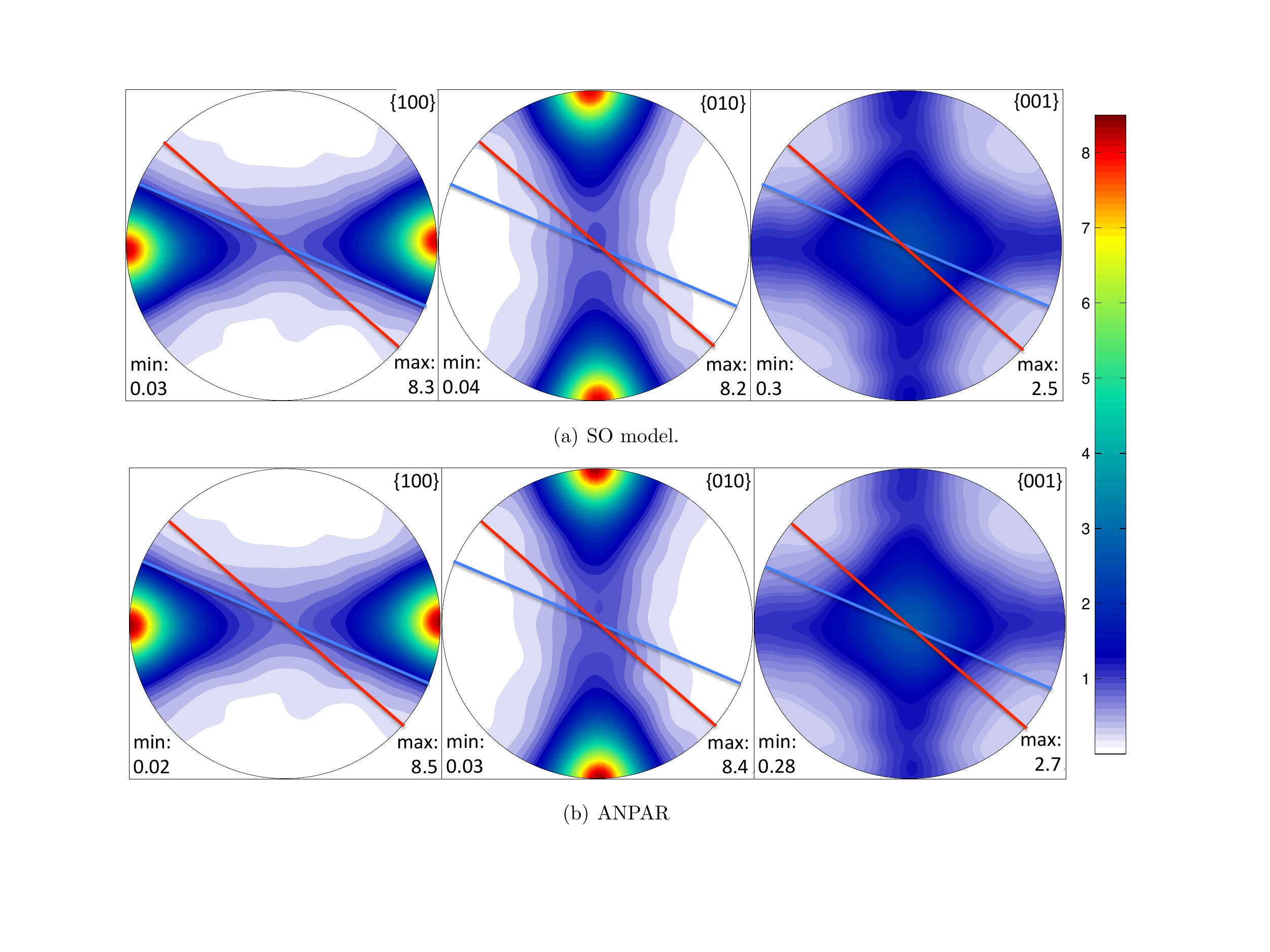}}
\end{center}
\caption{Same as Fig. \ref{fig_uniax}, but for the corner-flow model of a spreading ridge
shown in Fig. \ref{CF_geom}. Textures estimated by the SO and ANPAR models at the point of the filled FSE in Fig. \ref{CF_geom}. $\alpha=60^\circ$, $\varphi_0=10^\circ$, $\varphi=49^\circ$, $\chi=67^\circ$,
$r_{12} =1.047, r_{23} = -0.523$ ($r_0 = 0.6$), with CRSS ratios $\tau^{[1]}/\tau^{[2]} = 0.5$ and $\tau^{[2]}/\tau^{[3]} = 0.667$. The red and blue lines have the same meaning as in Fig. \ref{fig_sshear}.}\label{fig_CF}\end{figure}

Fig. \ref{fig_CF} shows the pole figures predicted by our theory
together with those predicted by the SO model for an equivalent strain $r_0=0.6$ ($r_{12}=1.047,$ $r_{23}=-0.523$). This was for the first streamline $\varphi_0=10^\circ$ in Fig. \ref{fig_CF}, with $\varphi=49^\circ$ and $\chi=67^\circ$. Again, the 
two sets of pole figures are almost identical, with a maximum difference in the predicted elasticity of
0.93 GPa. When comparing the two pole figures, the variance reduction is $99.0\%$. In this case, the speed of 
the ANPAR model is 5.8$\times 10^4$ greater than that of
the SO model (0.384 s for ANPAR vs. 2240 s for SO). 

\section{Discussion}

The new ANPAR method we describe in this article is both an accurate and computationally efficient alternative to existing methods for the simulation of CPO development in olivine. Benchmark 
tests against the second-order (SO) self-consistent model (Ponte-Caste\~neda, 2002) show that ANPAR runs
$2-8\times 10^4$ times faster, yet predicts textures that are nearly 
indistinguishable from those predicted by SO.

The ANPAR model has some similarities with the 
D-Rex model of Kaminski \& Ribe (2001).
In that model, the slip rates $\dot\gamma^{[s]}$ are predicted by
minimizing for each grain the misfit between the local and global
strain rate tensors. This \textit{ad hoc} principle yields 
\be
\dot\gamma^{[s]} = 2 A S^{[s]}_{ij} E_{ij}.
\label{gamdot}
\ee
where $A = 1$ if global strain compatibility is not enforced 
and $A=5$ if it is. Since the quantities $S^{[s]}_{ij}$ are
generalized spherical harmonics of degree 2, D-Rex
agrees with ANPAR concerning the spectral content
of the crystallographic spin. However, D-Rex assumes
that the spin does not depend on the background texture, and
so the amplitude $A$  does not
increase as strain accumulates. This is in contrast to the 
amplitudes $B$ and $C$ in ANPAR, both of which increase 
strongly with increasing strain in order to satisfy
global strain compatiblity. 

In constructing the ANPAR model we assumed that the spin tensor 
$A_{ijkl}$ can be written as the product of a tensor $Q_{ijkl}$
that depends only on the axial ratios $r_{12}$ and $r_{23}$ of the finite-strain ellipsoid
and a scalar $H$ that depends only on the relative slip system strengths
$p_{12}$ and $p_{23}$.
Although this seems to be a major assumption, the near-perfection 
of the fits we obtain to the SO predictions appears to justify it.

The simplicity of the ANPAR model is due in part
to the orthorhombic symmetry of olivine and the resulting  
orthogonality of the three dominant slip systems
(010)[100], (001)[100] and (010)[001]. The model can easily be extended to other 
orthorhombic minerals with less than 5 independent slip systems, such
as enstatite. However, the robust character of our results
leads us to suppose that our approach can be generalized  
to minerals with other symmetries and also to polyphase
rocks. Finally, the speed advantage of ANPAR over the SO
model holds out the possibility that it could be incorporated efficiently in 
3-D and time-dependent  simulations of mantle convection.

\section*{Acknowledgments}

The research leading to these results has received funding from the European Research Council under the European Union's Seventh Framework Program (FP7/2007-2013)/ERC grant agreement 240473 CoMITAC. We are grateful to A. Nowacki for helpful discussions.

\appendix
\label{app_amplitudes}
\section{Slip system-specific $\mathbf{S}$}

The Schmid tensor $\mathbf{S}$ is slip system-specific, and thus different for
each system. Let $S^{[s]}_{ij}$ be the Schmid tensor for slip system $s$, and write for simplicity
$S^{[1]}_{ij}=S_{ij}$.
For slip system $s=1$,  the relationships between the independent components of the Schmid tensor and the Eulerian angles are
\bsub
\label{BungeGSH}
\be
S_{11}-S_{22}=-\tfrac{1}{2}\left[\cos2\phi\sin2\psi\left(\cos^2\theta+1\right)+2\sin2\phi\cos\theta\cos2\psi\right],\ee
\be
S_{12}=-\tfrac{1}{4}\left[\sin2\phi\sin2\psi\left(\cos^2\theta+1\right)-2\cos2\phi\cos\theta\cos2\psi\right],\ee
\be
S_{13}=-\tfrac{1}{2}\sin\theta\left(\sin\phi\cos\theta\sin2\psi-\cos\phi\cos2\psi\right),\ee
\be
S_{23}=\tfrac{1}{2}\sin\theta\left(\cos\phi\cos\theta\sin2\psi+\sin\phi\cos2\psi\right),\ee
\be
S_{11}+S_{22}=-\tfrac{1}{2}\sin^2\theta\sin 2\psi.
\ee
\esub
Similarly, for $s=2$ we obtain
\bsub
\label{BungeGSH2}
\be
S^{[2]}_{11}-S^{[2]}_{22}=\sin\theta\left(\cos2\phi\cos\theta\sin\psi
+\sin2\phi\cos\psi\right),\ee
\be
S^{[2]}_{12}=\tfrac{1}{2}\sin\theta\left(\sin2\phi\cos\theta\sin\psi
-\cos2\phi\cos\psi\right),\ee
\be
S^{[2]}_{13}=-\tfrac{1}{2}\left(\sin\phi\cos2\theta\sin\psi
-\cos\phi\cos\theta\cos\psi\right),\ee
\be
S^{[2]}_{23}=\tfrac{1}{2}\left(\cos\phi\cos2\theta\sin\psi
+\sin\phi\cos\theta\cos\psi\right),\ee
\be
S^{[2]}_{11}+S^{[2]}_{22}=-\tfrac{1}{2}\sin2\theta\sin\psi.
\ee
\esub
Finally, for $s=3$ we find
\bsub
\label{BungeGSH3}
\be
S^{[3]}_{11}-S^{[3]}_{22}=\sin\theta\left(\cos2\phi\cos\theta\cos\psi
-\sin2\phi\sin\psi\right),\ee
\be
S^{[3]}_{12}=\tfrac{1}{2}\sin\theta\left(\sin2\phi\cos\theta\cos\psi
+\cos2\phi\sin\psi\right),\ee
\be
S^{[3]}_{13}=-\tfrac{1}{2}\left(\sin\phi\cos2\theta\cos\psi
+\cos\phi\cos\theta\sin\psi\right),\ee
\be
S^{[3]}_{23}=\tfrac{1}{2}\left(\cos\phi\cos2\theta\cos\psi
-\sin\phi\cos\theta\sin\psi\right),\ee
\be
S^{[3]}_{11}+S^{[3]}_{22}=-\tfrac{1}{2}\sin2\theta\cos\psi.
\ee
\esub

\section{Amplitude of the crystallographic rotation rate}

In this appendix we quantify the dependence of the slip
system amplitudes
$A^{[s]}_{ijkl}$ on the strains $r_{12}$ and $r_{23}$ and the
CRSS ratios $p_{12}$ and $p_{23}$. 
We assume that $\mathbf{A}$ can be expressed more compactly as
\be A^{[s]}_{ijkl}(r_{12},r_{23},p_{12},p_{23},m)=H^{[s]}(p_{12},p_{23},m)Q_{ijkl}(r_{12},r_{23})
\label{compactA}
\ee

\subsection{Limit $A_{ijkl}^{[s]}(r_{12}, r_{23}, 0, 0,3.5)$}

We first consider how the activities of slip systems with equal strengths
($p_{12} = p_{23} = 0$) vary as a function of strain. This enables us to find out how the $Q_{ijkl}$ depend on the parameters $r_{12}$ and $r_{23}$ that characterize the FSE. As we pointed out in Section 2.3, the value of $m=3.5$ is assumed for all slip systems. 

We explained in Section 4.2 the symmetries that allow us to reduce the number of independent non-zero components $Q_{ijkl}$ from 25 to just 2. The transformation rules (\ref{symrules}), derived 
by noting that the labelling of the coordinate axes is arbitrary, are given in their full form here:
\bsub
\begin{eqnarray}
Q_{1212}\left(r_{12},r_{23}\right)
&=&C\left(r_{12},r_{23}\right)=C\left(-r_{12},-r_{23}\right)
=C\left(r_{12},r_{31}\right)
=C\left(-r_{12},-r_{31}\right),\quad\quad\quad\\
Q_{1313}\left(r_{12},r_{23}\right)
&=&C\left(r_{31},r_{12}\right)
=C\left(-r_{31},-r_{12}\right)=C\left(r_{31},r_{23}\right)
=C\left(-r_{31},-r_{23}\right),\\
Q_{2323}\left(r_{12},r_{23}\right)
&=&C\left(r_{23},r_{31}\right)
=C\left(-r_{23},-r_{31}\right)=C\left(r_{23},r_{12}\right)
=C\left(-r_{23},-r_{12}\right),\\
Q_{1111}\left(r_{12},r_{23}\right)
&=&B\left(r_{12},r_{23}\right)=B\left(-r_{12},-r_{23}\right)
=B\left(r_{31},r_{23}\right)
=B\left(-r_{31},-r_{23}\right),\\
Q_{2222}\left(r_{12},r_{23}\right)
&=&B\left(r_{12},r_{31}\right)
=B\left(-r_{12},-r_{31}\right)=B\left(r_{23},r_{31}\right)
=B\left(-r_{23},-r_{31}\right),\\
B\left(r_{31},r_{12}\right)&=&B\left(-r_{31},-r_{12}\right)
=B\left(r_{23},r_{12}\right)
=B\left(-r_{23},-r_{12}\right)\nonumber\\
&=&
B\left(r_{12},r_{23}\right)
+2Q_{1122}\left(r_{12},r_{23}\right)
+Q_{2222}\left(r_{12},r_{23}\right)
\\
\therefore
Q_{1122}\left(r_{12},r_{23}\right)&=&\tfrac{1}{2}\left[B\left(r_{23},r_{12}\right)-B\left(r_{12},r_{23}\right)-B\left(r_{12},r_{31}\right)\right].
\end{eqnarray}
\label{Fullsym}
\esub
We also discovered, numerically, the symmetry condition
\be
\label{Asym}Q_{ijkl}\left(r_{12},r_{23}\right)
=Q_{ijkl}\left(-r_{12},-r_{23}\right).
\ee This means, for example, that the values of $Q_{ijkl}$ for uniaxial extension are identical to those for uniaxial compression.
Using (\ref{Asym}) and the complete symmetry transformations (\ref{Fullsym}), we can dramatically reduce the size of the parameter space $\left(r_{12},r_{23}\right)$ that we have to explore
numerically to determine how $B$ and $C$ depend on $r_{12}$ and $r_{23}$.

We determine values of $B$ and $C$ at sampled points in the $\left(r_{12},r_{23}\right)$-plane by fitting eqn (\ref{PSIDOT1}) (with $H^{[s]}(0,0)=1$) to the spin predicted by the SO model, using a standard least-squares procedure. These calculations yield the curves shown by solid lines in Figs. \ref{fig_A1111} and \ref{fig_A1212}. In each case the variance reduction $R>99.7\%$.
 
\subsubsection{Fitting functions.}

We fitted models to the $B$ and $C$ data, obtained by the above method. These fitting functions are fourth-order polynomials of the form
\be
\label{zmn}
\sum_{m=0}^4\sum_{n=0}^4z_{mn}r_{23}^m r_{12}^n,
\ee
where
\be z_{mn}=\begin{cases}\quad0\quad\textrm{if}\ m+n>4;\cr\quad0\quad\textrm{if}\ m+n \;\textrm{odd};\cr\quad z_{mn}\quad\textrm{otherwise.}\end{cases}\ee
The RMS errors of the fits to the $B$ and $C$ data are 0.039 and 0.0070, respectively. The values of the non-zero coefficients $z_{mn}$ are given in Table \ref{Table_coeffs}, for both fitting functions. The quality of these fits can be viewed in Figs 5 and 6.

\begin{table}
\begin{center}
\begin{tabular}{|c|c|c|c|c|c|c|c|c|c|}
\hline
&$z_{00}$&$z_{20}$&$z_{11}$&$z_{02}$&$z_{40}$&$z_{31}$&$z_{22}$&$z_{13}$&$z_{04}$\\\hline
$B$&2.241&0.3993&1.104&1.104&0.7619&2.507&5.518&6.023&3.012\\\hline
$C$&1.662&0.2046&0.1992&-0.7517&-0.01853&-0.02831&-0.4396&-0.4246&0.2085\\\hline
\end{tabular}
\end{center}
\caption{Values of the coefficients $z_{mn}$ in the fitting functions (\ref{zmn}), for $B\left(r_{12},r_{23}\right)$ and $C\left(r_{12},r_{23}\right)$.}\label{Table_coeffs}
\end{table}

\subsection{Limit $A_{ijkl}^{[s]}(0, 0, p_{12}, p_{23},3.5)$}

We next consider how the amplitudes depend on $p_{12}$ and $p_{23}$
at the initial instant ($r_0=r_{12} = r_{23} =0$) of the deformation. This allows us to discover how the functions $H^{[s]}$ in (\ref{compactA}) depend on the parameters that characterize the relative strength of the slip systems.  Here, we consider values of $p_{12}$ and $p_{23}$ in the range $[-\ln 4, \ln 4]\approx [-1.386, 1.386]$. We discovered numerically the following transformation rules:
\bsub
\begin{eqnarray}
H^{[1]}(p_{12},p_{23})&=&H^{[1]}(-p_{31},-p_{23})\\
H^{[2]}(p_{12},p_{23})&=&H^{[1]}(-p_{12},-p_{31})=H^{[1]}(p_{23},p_{31})\\
H^{[3]}(p_{12},p_{23})&=&H^{[1]}(-p_{23},-p_{12})=H^{[1]}(p_{31},p_{12})
\end{eqnarray}
\label{ptrans}
\esub
These transformations enable us not only to express $H^{[2]}$ and $H^{[3]}$ in terms of $H^{[1]}$, but also to reduce the size of the parameter space $\left(p_{12},p_{23}\right)$ that we have to explore
numerically to determine how $H^{[1]}$ depends on $p_{12}$ and $p_{23}$.

We now set $r_0=0$ and obtain $H^{[1]}$ data by calculating, via least-squares, the amplitudes $H^{[1]}$ that fit (\ref{PSIDOT1}) to 81 instantaneous numerical solutions of the SO
model for uniaxial compression, with equally spaced points in the $(p_{12},p_{23})$-plane with both $p_{12}$ and $p_{23}$ in the range
$[-1.386, 1.386]$. 
We then
fit quadratic polynomials, satisfying the above relations (\ref{ptrans}), to the collected $H^{[1]}$
data.
The results are
$$
H^{[1]}(p_{12},p_{23})=1-0.0295 p_{12} -0.0130 p_{23} - 0.00743p_{12}^2
$$
\begin{equation}
-0.00347p_{12} p_{23}-0.00333p_{23}^2.
\label{A1p}
\end{equation}
The RMS error of the fit is 0.0068. We apply the transformations (\ref{ptrans}) to (\ref{A1p}) to form analytical expressions for $H^{[2]}$ and $H^{[3]}$, respectively.

\subsection{General case}

A  general form for $A^{[s]}_{ijkl}(r_{12}, r_{23}, p_{12}, p_{23})$,
that is consistent with the above limiting cases, is obtained by substituting the models (\ref{zmn}) and (\ref{A1p}) into (\ref{compactA}).
The resulting expression is the one we  
use in all our numerical calculations.

We have fitted this model to the spin predicted by the SO model for random background textures (formed from various $r_{12}$, $r_{23}$, $p_{12}$ and $p_{23}$ values) and random instantaneous SRT's. Remarkably, in each case, the variance reduction $R> 99.1\%$ and in most cases $R>99.7\%$.

\section{Variance reduction between pole figures}

To calculate the variance reduction between the pole figures shown in Section 5,
we use the transformations
\be
\phi^\ast_n=\frac{2}{\pi}\phi_n,\quad
\theta^\ast_n=1 + \cos \theta_n,\quad
\psi^\ast_n=\frac{2}{\pi}\psi_n.
\ee
to map the Euler angles $\left(\phi_n,\theta_n,\psi_n\right)$ of each grain onto an `Euler cube'
$\left(\phi_n^{\ast},\theta_n^{\ast},\psi_n^{\ast}\right)$. The Euler cube has a uniform metric, and each of its sides is of length 2.0. 
Let the set of Euler angles for each grain for the SO and ANPAR approximations be denoted by $\left(\phi^\ast_{(n,SO)},\theta^\ast_{(n,SO)},\psi^\ast_{(n,SO)}\right)$ and $\left(\phi^\ast_{(n,AN)},\theta^\ast_{(n,AN)},\psi^\ast_{(n,AN)}\right)$, respectively. For each grain the `distance', $d_n$, between these two sets of Euler angles is calculated by
\be
d_n=\sqrt{\left(\phi^\ast_{(n,SO)}-\phi^\ast_{(n,AN)}\right)^2
+\left(\theta^\ast_{(n,SO)}-\theta^\ast_{(n,AN)}\right)^2
+\left(\psi^\ast_{(n,SO)}-\psi^\ast_{(n,AN)}\right)^2}
\ee
The variance reduction between the two pole figures is then given by
\be
R=1-\frac{\sum_{n=1}^N d_n^2}{\sum_{n=1}^N\left[\left(\phi^\ast_{(n,SO)}\right)^2
+\left(\theta^\ast_{(n,SO)}\right)^2
+\left(\psi^\ast_{(n,SO)}\right)^2\right]},
\ee

\label{lastpage}


\begin{thebibliography}{}

\bibitem{} Ammann, M. W., Walker, A. M., Stackhouse, S., Forte, A. M., Wookey, J., Brodholt, J. P., Dobson, D. P., 2014. Variation of thermal conductivity and heat flux at the Earth's core mantle boundary, \textit{Earth Planet. Sci. Lett.},  \textbf{390}, 175--185.

\bibitem{} Bachmann, F., Hielscher, R., Schaeben, H., 2010. Texture analysis with MTEX - free and open source software toolbox. \textit{Solid State Phenom.}, \textbf{160}, 63--68.

\bibitem{} Bai, Q., Mackwell, S.J. \& Kohlstedt, D.L., 1991. High-temperature creep of olivine single crystals. 1. Mechanical results for buffered samples, \textit{J. Geophys. Res.}, \textbf{96}, 2441–2463.

\bibitem{}
Batchelor, G. K., 1967. An Introduction to Fluid Mechanics, \textit{Cambridge Univ. Press, New York.}

\bibitem{} Becker, T., Kustowski, B., Ekstr\"om, G. 2008. Radial seismic anisotropy as a constraint for upper mantle rheology, \textit{Earth Planet. Sci. Lett.}, \textbf{267}, 213--227.

\bibitem{}
Becker, T., Lebedev, S., Long, M., 2012. On the relationship between azimuthal anisotropy from shear wave splitting and surface wave tomography. \textit{J. Geophys. Res.}, \textbf{117}, B01306.

\bibitem{}
Bonnin, M., Tommasi, A., Hassani, R., Chevrot, S., Wookey, J., Barruol, G., 2012. Numerical modelling of the upper-mantle anisotropy beneath a migrating strike-slip plate boundary: the San Andreas Fault system. \textit{Geophys. J. Int.}, \textbf{191}, 436--458.


\bibitem{} Bunge, H. J. 1982. Texture Analysis in Materials Science. \textit{Butterworths, London.}

\bibitem{}
Castelnau, O., Blackman, D. K., Lebensohn, R. A. \& Castaneda, P. P., 2008.
Micromechanical modeling of the viscoplastic behavior of olivine,
\textit{J. Geophys. Res.}, \textbf{113}, B09202,

\bibitem{}
Castelnau, O., Blackman, D. K. \& Becker, T.W.,  2009.
Numerical simulations of texture development and associated rheological anisotropy
in regions of complex mantle flow, \textit{Geophys. Res. Lett.},
\textbf{36}, L12304, doi:10.1029/2009GL038027.

\bibitem{}
Castelnau, O., Cordier, P., Lebensohn, R., Merkel, S., Raterron, P., 2010. Microstructures and rheology of the Earth's upper mantle inferred from a multiscale approach. \textit{Comptes Rendus Physique}, \textbf{11}, 304-–315.

\bibitem{}
Clement, A., 1982.
Prediction of deformation texture using a physical principle of conservation.
\textit{Mater. Sci. Eng.}, \textbf{55}, 203--210.

\bibitem{}
Conder, J. A \& Wiens, D. A., 2007.
Rapid mantle flow beneath the Tonga volcanic arc,
\textit{Earth Planet. Sci. Lett.}, \textbf{264}, 299--307.

\bibitem{}
Cottaar, S., Li, M., McNamara, A. K., Romanowicz, B. Wenk, H.-R., 2014. Synthetic seismic anisotropy models within a slab impinging on the core–mantle boundary. \textit{Geophys. J. Int.}, \textbf{199}, 164--177.

\bibitem{}
Crampin, S., 1984. An introduction to wave-propagation in anisotropic media. \textit{Geophys. J. Roy. Astr. S.}, \textbf{76}, 17–-28.

\bibitem{}
Di Leo, J. F., Walker, A. M., Li, Z.-J., Wookey, J., Ribe, N. M., Kendall, J.-M., Tommasi, A., 2014.
Development of texture and seismic anisotropy during the onset of subduction.
\textit{Geochem. Geophys.
Geosyst.}, \textbf{15}, 192--212.

\bibitem{} Dobson, D. P., Miyajima, N., Nestola, F., Alvaro, M., Casati, N., Liebske, C., Wood, I. G., Walker, A. M. 2013. Strong inheritance of texture between perovskite and post-perovskite in the D$^{\prime\prime}$ layer, \textit{Nat. Geosci.}, doi: 10.1038/NGEO1844

\bibitem{}Faccenda, M. \& Capitanio, F. A., 2012. Development of mantle seismic anisotropy during subduction-induced 3-D flow. \textit{Geophys. Res. Lett.}, \textbf{39}, L11305. doi: 10.1029/2012GL051988

\bibitem{}
Faccenda, M. \& Capitanio, F. A., 2013. Seismic anisotropy around subduction zones: Insights from three-dimensional modeling of upper mantle deformation and SKS splitting calculations. \textit{Geochem. Geophys.
Geosyst., in press}, doi: 10.1029/2012GC004451

\bibitem{}
Faccenda, M., 2014. Mid mantle seismic anisotropy around subduction zones. \textit{Phys. Earth Planet. In.}, \textbf{227}, 1--19.

\bibitem{}
Grennerat, F., Montagnat, M., Castelnau, O., Vacher, P., Moulinec, H., Suquet, P., Duval, P, 2012. Experimental characterization of the intragranular strain field in columnar ice during transient creep. \textit{Acta Mater.} \textbf{60}, 3655--3666.

\bibitem{}
Kaminski, E. \& Ribe, N.,  2001. A kinematic model for recrystallization
and texture development in olivine polycrystals,
\textit{Earth Planet. Sci. Lett.}, \textbf{189}, 253--267.

\bibitem{}
Kaminski, E. \& Ribe, N. M.,  2002. Time scales for the evolution of
seismic anisotropy in mantle flow, \textit{Geochem. Geophys.
Geosyst.}, \textbf{10}, doi:10.1029/2001GC000222.

\bibitem{}
Kaminski, E., Ribe, N. M., Browaeys, J. T., 2004.
D-Rex, a program for calculation of seismic anisotropy in the convective
upper mantle, \textit{Geophys. J. Int.}, \textbf{158}, 744--752.

\bibitem{}
Kanit, T., Forest, S., Galliet, I.,  Mounoury, V., and Jeulin, D., 2003. Determination of the size of the representative volume element for random composites: Statistical and numerical approach, \textit{Int. J. Solids Struct.}, \textbf{40}, 3647--3679.

\bibitem{}
Kellogg, L. H. \& Turcotte, D. L., 1990. Mixing and the distribution of 
heterogeneities in a chaotically convecting mantle,
\textit{J. Geophys. Res.} \textbf{95}, 421--432.

\bibitem{}
Lachenbruch, A. H. \& Nathenson, M., 1974. Rise of a variable viscosity fluid in a steadily spreading wedge-shaped conduit with accreting walls. \textit{U. S. Geol. Surv. Open File Rep.} 74--251.

\bibitem{}
Lassak, T. M., Fouch, M. J., Hall, C. E. \& Kaminski, E., 2006.
Seismic characterization of mantle flow in subduction systems:
Can we resolve a hydrated mantle wedge?,
\textit{Earth Planet. Sci. Lett.}, \textbf{243}, 632--649.

\bibitem{}
Lebensohn, R.A., Tom\'e, C.N., 1993.
 A selfconsistent approach for the simulation of plastic deformation and texture development of polycrystals: application to zirconium alloys. 
\textit{Acta Metall. Mater.}, \textbf{41}, 2611--2624.

\bibitem{}
Lebensohn, R. A., 2001. N-site modeling of a 3D viscoplastic polycrystal using fast Fourier transform, \textit{Acta Mater.}, \textbf{49}, 2723--2737.

\bibitem{}
Lebensohn, R. A., Rollett, A., Suquet, P., 2011. Fast Fourier Transform-based modelling for the determination of micromechanical fields in polycrystals. \textit{JOM} \textbf{63}, 13--18.

\bibitem{}
Lev, E. \& Hager, B. H., 2008.
Prediction of anisotropy from flow models: A comparison of three methods,
\textit{Geochem. Geophys.
Geosyst.}, \textbf{9}, Q07014, doi:10.1029/2001GC000222.

\bibitem{} Long, M. D. 2013. Constraints on subduction geodynamics from seismic anisotropy, \textit{Rev. Geophys.}, \textbf{51}, 76--112.

\bibitem{} Long, M. D. \& Becker, T. W. 2010. Mantle dynamics and seismic anisotropy, \textit{Earth Planet. Sci. Lett.}, \textbf{297}, 341--354.

\bibitem{}
Mainprice, D., Tommasi, A., Couvy, H., Cordier, P., Frost, D., 2005. Pressure sensitivity of olivine slip systems and seismic anisotropy of Earth’s upper mantle. \textit{Nature}, \textbf{433}, 731--733.

\bibitem{}
Mainprice, D., Tommasi, A., Ferr\'e, D., Carrez, P., Cordier, P., 2008.
Predicted glide systems and crystal preferred orientations of polycrystalline silicate {M}g-Perovskite at high pressure: {I}mplications for the seismic anisotropy in the lower mantle.
\textit{Earth Planet. Sci. Lett.}, \textbf{271}, 135--144.

\bibitem{}
Masson, R., Bornert, M., Suquet, P., Zaoui, A., 2000. 
An affine formulation for the prediction of the effective
properties of nonlinear composites and polycrystals.
\textit{J. Mech. Phys. Solids}, \textbf{48}, 1203--1227.

\bibitem{}
McKenzie, D., 1979. Finite deformation during fluid flow, \textit{Geophys. J. R. Astron. Soc.}, \textbf{58}, 689--715.

\bibitem{}
Merkel, S., McNamara, A., Kubo, A., Speziale, S., Miyagi, L., Meng, Y., Duffy, T., Wenk, H.-R, 2007. Deformation of (Mg,Fe)SiO3 post-perovskite and D” anisotropy. \textit{Science}, \textbf{316}, 1729--1732.


\bibitem{}
Molinari, A., Canova, G.R., Ahzi, S., 1987. 
A selfconsistent approach of the large deformation polycrystal
viscoplasticity. \textit{Acta Metall.}, \textbf{35}, 2983--2994.

\bibitem{}
Montagner, J.-P. \& Tanimoto, T., 1991. Global upper mantle tomography of seismic velocities and anisotropies. \textit{J. Geophys. Res-Sol. Ea.}, \textbf{96}, 20337--20351.

\bibitem{}
Moulinec, H., Suquet, P., 1998. A numerical method for computing the
overall response of nonlinear composites with complex microstructure,
\textit{Comput. Methods Appl. Mech. Eng.}, \textbf{157}, 69--94.

\bibitem{}
Muhlhaus, H., Moresi, L., Cada, M., 2004. Emergent Anisotropy and Flow Alignment in Viscous Rock. \textit{Pure Appl. Geophys}. \textbf{161}, 2451--2463. doi: 10.1007/s00024-004-2575-5.

\bibitem{}
Nicolas, A., \& Christensen, N., 1987. Formation of anisotropy in upper mantle peridotites: A review, in Composition, Structure and Dynamics of the Lithosphere-Asthenosphere System. \textit{Geodyn. Ser.}, \textbf{16}, 111--123. doi:10.1029/GD016p0111.

\bibitem{} Nowacki, A., Walker, A. M., Wookey, J., Kendall, J.-M. 2013. Evaluating post-perovskite as a cause of D$^{\prime\prime}$ anisotropy in regions of palaeosubduction. \textit{Geophys. J. Int.}, \textbf{192}, 1085--1090.

\bibitem{}
Ponte Casta\~neda, P., 2002.
Second-order homogenization estimates for nonlinear composites incorporating field fluctuations. I. Theory, and II. Applications.
\textit{J. Mech. Phys. Solids}, \textbf{50}, 737--782.

\bibitem{}
Raterron, P., Detrez, F.,  Castelnau, O., Bollinger, C. Cordier, P., Merkel., S., 2014. Multiscale modeling of upper mantle plasticity: From single-crystal rheology to multiphase aggregate deformation. \textit{Phys. Earth Planet. In.}, \textbf{228},
232--243.

\bibitem{}
Ribe, N. M, 1989. Seismic anisotropy and mantle flow, \textit{J. Geophys.
Res.}, \textbf{94}, 4213--4223.

\bibitem{}
Ribe, N. M. \& Yu, Y., 1991. A theory for plastic deformation and
textural evolution of olivine polycrystals, \textit{J. Geophys. Res.}, \textbf{96},
8325--8335.

\bibitem{}
Ribe, N. M, 1992. On the relation between seismic anisotropy and
finite strain, \textit{J. Geophys.
Res.}, \textbf{97}, 8737--8747.

\bibitem{}
Sarma, G. B. \& Dawson, P. R., 1996. Effects of interactions among crystals on the inhomogeneous deformation of polycrystals, \textit{Acta Mater.}, \textbf{44}(5), 1937-1953.

\bibitem{}
Silver, P. \& Chan, W., 1991. Shear-wave splitting and subcontinental mantle deformation. \textit{J. Geophys. Res.}, \textbf{96}, 16429--16454.

\bibitem{}
Silver, P., 1996. Seismic anisotropy beneath the continents: Probing the depths of geology. \textit{Annu. Rev. Earth Planet. Sci.}, \textbf{24}, 385--432.

\bibitem{}
Suquet, P., Moulinec, H., Castelnau, O., Montagnat, M., Lahellec, N., Grennerat, F., Duval, P., Brenner, R., 2012. Multiscale modeling of the mechanical behavior of polycrystalline ice under transient creep, \textit{Procedia IUTAM}, \textbf{3}, p. 64--78.

\bibitem{}
Tommasi, A., Tikoff, B., Vauchez, A., 1999. Upper mantle tectonics: three-dimensional deformation, olivine crystallographic fabrics and seismic properties. \textit{Earth Planet. Sc. Lett.}, \textbf{168}, 173--186.

\bibitem{}
Tommasi, A., Mainprice, D., Canova, G., Chastel, Y., 2000.
Viscoplastic self-consistent and equilibrium-based modeling of olivine lattice preferred orientations: {I}mplications for the upper mantle seismic anisotropy.
\textit{J. Geophys. Res.}, \textbf{105},
7893--7980.

\bibitem{}
Tommasi, A., Mainprice, D., Cordier, P., Thoraval, C., Couvy, H., 2004. Strain-induced seismic anisotropy of wadsleyite polycrystals and flow patterns in the mantle transition zone. \textit{J. Geophys. Res.}, \textbf{109}. doi: 10.1029/2004JB003158

\bibitem{}
Tommasi, A., Knoll, M.,  Vauchez, A., Signorelli, J., Thoraval, C., Log\'e, R., 2009. Structural reactivation in plate tectonics controlled by olivine crystal anisotropy. \textit{Nat. Geosci.}, \textbf{2}, 423--427.

\bibitem{}
Tovish, A., Schubert, G., Luyendyk, B. P., 1978. Mantle flow pressure and the angle of subduction: Non-Newtonian corner flows. \textit{J. Geophys. Res.}, \textbf{83}, 5892--5898.

\bibitem{}
Walker, A. M., Forte, A. M., Wookey, J., Nowacki, A., Kendall, J.-M., 2011.
Elastic anisotropy of {D}$^{\prime\prime}$ predicted from global models of mantle flow.
\textit{Geochem. Geophys.
Geosyst.}, \textbf{12}, Q07014, doi:10.1029/2011GC003732.

\bibitem{}
Walker, A. M. \& Wookey, J., 2012. MSAT - A new toolkit for the analysis
of elastic and seismic anisotropy. \textit{Comput. Geosci.}, \textbf{49},
81-90. doi:10.1016/j.cageo.2012.05.031


\bibitem{}
Wenk, H.-R., Bennett, K., Canova, G. R., Molinari, A., 1991.
Modelling plastic deformation of peridotite with the self-consistent theory.
\textit{J. Geophys. Res.}, \textbf{96},
8337--8349.

\bibitem{}
Wenk, H.-R. \& Tom\'e, C. N., 1999.
Modeling dynamic recrystallization of olivine aggregates deformed in simple shear.
\textit{J. Geophys. Res.}, \textbf{104},
25513--25527.

\bibitem{}
Wenk, H.-R., Speziale, S., McNamara, A., Garnero, E., 2006.
Modeling lower mantle anisotropy development in a subducting slab.
\textit{Earth Planet. Sci. Lett.}, \textbf{245},
302--314.



\bibitem{}
Wenk, H.-R., Cottaar, S., Tom\'e, C. N., McNamara, A., Romanowicz, B., 2011.
Deformation in the lowermost mantle: From polycrystal plasticity to seismic anisotropy.
\textit{Earth Planet. Sci. Lett.}, \textbf{306},
33--45.

\end{thebibliography}
\end{document}